\documentclass[a4paper,10pt]{article}
\usepackage{authblk}

\usepackage[english]{babel}
\usepackage[utf8]{inputenc}
\usepackage{csquotes}
\usepackage{amsfonts,amsmath,amsthm}
\usepackage{empheq}
\usepackage{cancel}

\usepackage{bbm}
\usepackage{tikz}
\usetikzlibrary{spy} 

\usepackage[titletoc,title]{appendix}
\usepackage[backend=bibtex8,doi=false,eprint=false,giveninits=true,isbn=false,style=numeric-comp,url=false,maxnames=99]{biblatex}
\makeatletter
\def\blx@maxline{77}
\makeatother
\DeclareRedundantLanguages{english,german,french}{english,german,ngerman,french}

\usepackage{cases}
\usepackage{mathabx}
\usepackage{xfrac}
\usepackage{fancyhdr}
\usepackage{color}
\usepackage[colorlinks]{hyperref}
\definecolor{blue75}{rgb}{0,0,.75}
\definecolor{green75}{rgb}{0,.75,0}

\newcommand{\cred}{\color{red}}

\newcommand{\eps}{\varepsilon}
\newcommand{\pt}{{\partial}}
\newcommand{\uu}{{\bf u}}

\hypersetup{colorlinks=true, urlcolor=blue75,linkcolor=blue75,citecolor=green75,pdfstartview=FitB,bookmarksopen=true,bookmarksopenlevel=1}
\usepackage[a4paper, left=2.5cm, right=2.5cm, top=2.5cm,bottom=2cm]{geometry}
\usepackage{constants}
\newconstantfamily{C}{
symbol=C,
format=\parenthezises,
reset={section}
}
\usepackage{enumerate}

\usepackage{graphicx}
\graphicspath{ {images/} }
\usepackage{wrapfig}
\usepackage{figbib}
\usepackage{caption}
 \usepackage{subcaption}
\usepackage{changes}
\allowdisplaybreaks
\begin{document}
\newcommand{\R}{\mathbb{R}}
\newcommand{\N}{\mathbb{N}}
\newcommand{\Ss}{\mathbb{S}}

\newcommand{\calL}{\mathcal{L}}
\newcommand{\calI}{\mathcal{I}}
\newcommand{\Ker}{\text{Ker }}
\newcommand{\spn}{\text{span }}

\newcommand{\port}{p^\perp}

\newcommand{\ua}{u^{\alpha}}
\newcommand{\ub}{u^{\beta}}
\newcommand{\uaq}{u^{q-1+\alpha}}
\newcommand{\Tm}{T_{max}}
\newcommand{\oa}{\bar{\Omega}}
\newcommand{\bu}{\bar{u}}
\newcommand{\ot}{\Omega \times (0,T)}
\newcommand{\uj}{u_{{\cred l}k}}
\newcommand{\ujk}{u_{{\cred l}m_{\cred o}}}
\newcommand{\hj}{h_{{\cred l}k}}
\newcommand{\hjk}{h_{{\cred l}km_{\cred o}}}
\newcommand{\ck}{C^{\kappa,\frac{\kappa}{2}}(\oa \times [0,T])}
\newcommand{\io}{\int_{\Omega}}
\newcommand{\tr}{\text{tr}}
\newcommand{\D}{\mathbb{D}}
\newcommand{\dij}{d_{ij}}
\newcommand{\td}{\text{d}}
\newcommand{\bh}{\bar{h}}
\newcommand{\Om}{\Omega }
\newcommand{\Sbb}{\mathbb S^{n-1}}

\newcommand{\etaeta}{{\bf \eta}}
\newcommand{\nn}{{\mathbf{n}}}

\newtheorem{Theorem}{Theorem}[section]
\newtheorem{Assumptions}[Theorem]{Assumptions}
\newtheorem{Corollary}[Theorem]{corollary}
\newtheorem{Convention}[Theorem]{convention}
\newtheorem{Definition}[Theorem]{Definition}

\newtheorem{Lemma}[Theorem]{Lemma}
\newtheorem{Notation}[Theorem]{Notation}
\newtheorem{Remark}[Theorem]{Remark}
\theoremstyle{definition}
\newtheorem{Example}[Theorem]{Example}
\numberwithin{equation}{section}
\title{An in-silico approach to meniscus tissue regeneration: Modeling, numerical simulation, and 
experimental analysis}
\author{Elise Grosjean, Alex Keilmann, Henry J\"ager, 
	Shimi Mohanan, 
	 Claudia Redenbach, Bernd Simeon, and Christina Surulescu\thanks{\href{mailto:surulescu@mathematik.uni-kl.de}{surulescu@mathematik.uni-kl.de}} \\
	RPTU Kaiserslautern-Landau, Department of Mathematics,\\ Postfach 3049, 67653 Kaiserslautern, 
	Germany
	 \and 
	Luisa de Roy, Graciosa Teixera, and Andreas Martin Seitz\\
	Institute of Orthopaedic Research and Biomechanics,\\ Centre of Musculoskeletal Research Ulm,\\
	University Ulm, Helmholtzstraße 14, 89081 Ulm, Germany \and
	Martin Dauner, Carsten Linti, and G\"unter Schmidt\\
	Deutsche Institute f\"ur Textil- und Faserforschung (DITF)\\
	Koerschtalstr. 26, 73770 Denkendorf}

\date{\today}
\maketitle

\begin{abstract} 
	\noindent
We develop a model the dynamics of human mesenchymal
stem cells (hMSCs) and chondrocytes evolving in a nonwoven polyethylene terephtalate (PET) scaffold impregnated with hyaluron and supplied with a differentiation medium. The scaffold and the cells are assumed to be contained in a bioreactor with fluid perfusion. The differentiation of hMSCs into chondrocytes favors the production of extracellular matrix (ECM) and is influenced by fluid stress. The model
takes deformations of ECM and PET 
scaffold into account. The scaffold structure is explicitly included by statistical assessment of the fibre distribution from CT images. The effective macroscopic equations  are obtained by appropriate upscaling from
dynamics on lower (microscopic and mesoscopic) scales and feature in the motility terms an explicit cell diffusion tensor encoding the assessed anisotropic scaffold structure. Numerical simulations show its influence on the overall cell and tissue dynamics.
\end{abstract}

\section{Introduction}\label{sec1}

During the last decades, mathematical modeling and simulation have become valuable tools for investigating complex biomedical systems. They contribute significantly to understanding different aspects of a biological process, often allowing to extend the study to related, mutually conditioned processes. In this spirit. the present paper is concerned with modeling, simulation and experimental validation for a prominent biomedical problem, the meniscus regeneration and involved cell and tissue-level phenomena. \\[-2ex]

\noindent
Clinical studies indicate that partial and total meniscectomies lead to prevalence of premature osteoarthritis in knee joints. Therefore, substantial efforts are being made towards finding adequate regenerative tissue for meniscus replacement. Although there are some approaches and even commercially available products, to date the optimal substitute has not been developed. Most regenerative approaches are clinically motivated and focus rather on the practical application than on the micro- and macroscopic cellular mechanisms and the interactions with the scaffold material. The latter viewpoint is promising in the sense that it aims to understand the basic control mechanisms in cell-scaffold interactions under different environmental parameters, thus providing a selective prognosis of the most conducive combinations of these parameters.\\[-2ex]

\noindent
With respect to the in-silico modeling and simulation, a major challenge lies in the well-posed and numerically efficient coupling of the processes at the cell level with the macroscopic behavior and the mechanical properties of the tissue. The active processes at the cell level, such as cell differentiation and matrix synthesis, have a strong impact on the resulting tissue structure, while macroscopic effects in turn are important stimuli for the processes at the microscopic level. Moreover, the time scales of the different processes differ vastly.\\[-2ex]

\noindent
A key feature of our work is the use of accompanying experiments that are based on a nonwoven scaffold in a
perfusion chamber, allowing in-vitro investigations for the development of chondrocytes and adipose tissue derived stem cells. In this framework, crucial stimuli to achieve relevant proliferation, migration and differentiation can be identified by state-of-the-art measurements. While meaningful clinical data is very difficult to obtain from the interior meniscus tissue, this off-the-wall approach provides comprehensive underpinnings for the mathematical modeling and numerical simulation.\\[-2ex]

\noindent
We give next a short overview on related work and refer to \cite{WSE-H} for a recent  review of mathematical modeling in tissue regeneration in a larger sense. 
In spite of the increasing interest attached to meniscus tissue engineering, there are relatively few mathematical models for the dynamics of  involved processes. Essential aspects concern degradation of engineered fibers, migration/differentiation of stem cells into/within the scaffold, and production of tissue by  chondrocytes. Thereby, 
stem cell (de)differentiation seems to play a major role. It can be triggered by mechanical stress \cite{altman2002cell,park2013control}, tissue stiffness \cite{mousavi2015role}, topography of the scaffold \cite{ghasemi2015structural}, or by chemical cues present in the extracellular space \cite{freymann2013toward,mauck2007regional,mishima2008chemotaxis}.\\[-2ex]

\noindent 
Continuous settings have the advantage of enabling mathematically well-founded qualitative  analyses and fast, efficient numerical simulations. This usually makes up for them  typically including less details than their discrete or hybrid counterparts. The vast majority of continuous approaches involve reaction-diffusion-(transport) equations (RD(T)Es) to describe the dynamics of \textit{macroscopic} cell densities interacting with soluble or insoluble extracellular components and leading to tactic cell migration.  There is a vast literature concerning such systems; we refer to the reviews in \cite{bellomo2015toward} and \cite{kolbe2021modeling}, and e.g., to \cite{stinner2016global,stinner2014global,zhigun2018strongly,zhigun2016global} for the analysis of models paying enhanced attention to effects of heterogeneous  tissues on cell migratory and proliferative behavior. Nonlocal RDTE models also attach importance to phenomena taking place at longer range than just at the very location of a cell; this is particularly relevant for high cell densities and for more detailed descriptions cell-cell and/or cell-tissue interactions. We refer to \cite{CPSZ} for a review of such models; in \cite{eckardt2020nonlocal} we established rigorous relationships with cell-cell and cell-tissue  adhesion models involving spatial nonlocality. RDTE models for tissue regeneration taking into account the evolution of cell populations with various phenotypes are rather rare; some of them \cite{geris2008angiogenesis,BAILONPLAZA2001,Campbell2019,Campbell2019-b} account for biochemical influences, others also for mechanical ones see e.g. \cite{andreykiv2008simulation,GmezBenito2005,Pohlmeyer2013,Ribeiro2015}. We are, however, not aware  of any formulations capturing in this context the topography of underlying tissue - the less so, as far as meniscus regeneration is concerned. The works \cite{Barocas1994,Barocas1997} provide a rather detailed description of anisotropic tissue-like structures populated with cells whose migratory behavior is influenced by such surroundings and their mechanics, but in the setting of a biphasic theory and the fibre orientation distribution is thereby not assessed statistically from experiments.\\[-2ex]

\noindent
Multiphase models are another continuous approach, where cell populations are seen as components of a mixture also containing fluid(s) in which chemical cues are dissolved  and/or tissue. Mass and momentum balance are typically required for each of the involved phases, supplemented with appropriate closure laws, see e.g. \cite{AP08,barocas1997anisotropic,Lemon2006,lemon2007travelling}; the review \cite{klika2016overview} of multiphase cartilage mechanical modeling explicitly excludes descriptions of cell behavior involved in the process. The advantage of mixture-based approaches is their ability to pay enhanced  attention to biomechanics (\cite{barocas1997anisotropic}), however their rigorous mathematical analysis is challenging and has scarcely been addressed, mainly just in 1D, and for substantially simplified settings leading to smaller RDTE systems (see e.g., \cite{Evje-Winkler20} and references therein). Connections between multiphase models and RDTE systems for (tumor) cell migration and spread in the extracellular matrix have been done in \cite{jackson2002mechanical} 
(1D case) and \cite{kumar2021multiphase} (higher dimensions, more complex setting). Multiphase models are commonly macroscopic, however some works within this framework also addressed multiscality \cite{HOLDEN2019,ODea2014}, see also the review \cite{Borgiani2017} and references therein.\\[-2ex]

\noindent
\textit{Mesoscale} models involve cell density distributions depending on time, position, cell velocity, and so-called activity variables in the kinetic theory of active particles framework \cite{bellomo2017quest}. They are intermediates between single cell and macroscopic dynamics and feature kinetic transport equations (KTEs), with integral terms describing velocity innovations as consequences of cell reorientations, along with chemo- and haptotactic bias. Such models are able to incorporate detailed information about the anisotropy of fibrous tissue surrounding the cells and capture effects of such topography on the motility and phenotypic switch of cellular matter. Adequate upscaling and moment closure procedures lead again, for the cell populations of interest, to macroscopic RDTEs carrying in their motility (myopic diffusion and taxis) and source terms (proliferation/decay/phenotypic switch) the said influences of tissue structure and chemoattractants and/or -repellents. A variety of such models have been developed and investigated \cite{conte2023mathematical,corbin2021modeling,dietrich2022multiscale,engwer2015glioma,engwer2016effective,kumar2021multiscale,painter2013mathematical,zhigun2022novel}, primarily in connection to glioma invasion in brain tissue, where patient-specific anisotropy seems to be essential for the spread of cancer cells. Thereby, the recent works \cite{corbin2021modeling,dietrich2022multiscale,zhigun2022novel} also proposed some ways to include biomechanics in the description of cell and tissue dynamics. 
The cell motility terms of the obtained macroscopic PDEs involve flux-limited diffusion and taxis. While the vast majority of passages from KTEs to RDTEs is informal, we introduced in \cite{zhigun2022novel} a novel method to derive a rigorous macroscopic limit.
All mentioned model types can accommodate multiscality  either through adequate couplings of ODEs with PDEs as e.g., in \cite{Lorenz2014,stinner2014global}, or by some 
of the kinetic variables being obtained on another modeling level, as in \cite{corbin2021modeling,dietrich2022multiscale,engwer2015glioma,engwer2016effective,kumar2021multiscale}.\\[-2ex]

\noindent
The paper is organized as follows: Section \ref{sec:modeling} is dedicated to obtaining the macroscopic model comprising effective RDTEs for the dynamics of involved cell populations (stem cells and their differentiated counterparts), which are then coupled to fluid mechanics and deformations of scaffold and newly generated tissue, but also to the equations for the evolution of hyaluron (which impregnates the scaffold), differentiation factor, and newly formed tissue. Section \ref{sec:experiments} provides information about the performed experiments and the data processing. In Section \ref{sec:numerics} we perform numerical simulations of the model. Finally, Section \ref{sec:discussion} contains a discussion of the results and an outlook.

\section{Mathematical modeling}\label{sec:modeling}
\begin{figure}
	\begin{subfigure}{\textwidth}
		\centering
		\includegraphics[width=\textwidth]{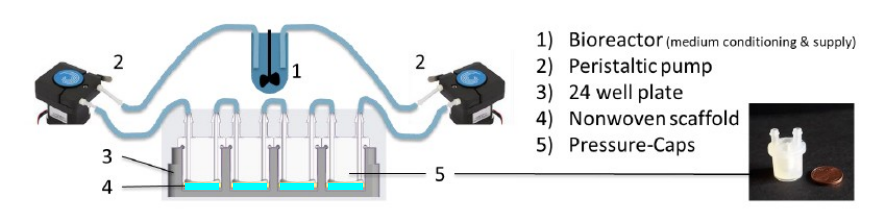}
		\caption{Setup for array of perfusions chambers (pressure caps)}
	\end{subfigure}
	\begin{subfigure}{\textwidth}
		\centering
		\includegraphics[width=0.75\textwidth]{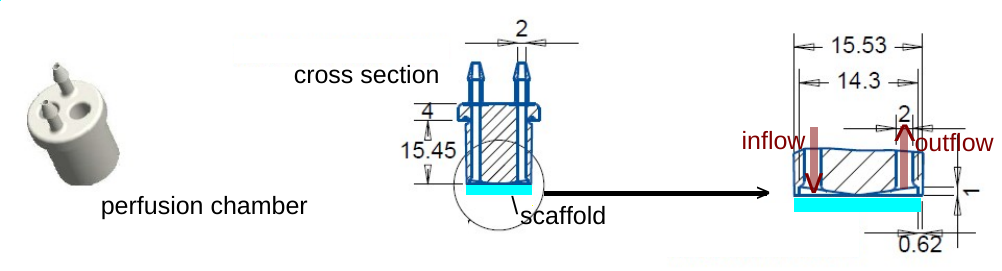}
		\caption{Geometry of the perfusion chamber (lengths in [mm])}
	\end{subfigure}
	\begin{subfigure}{\textwidth}
		\centering
		\includegraphics[width=\textwidth/2]{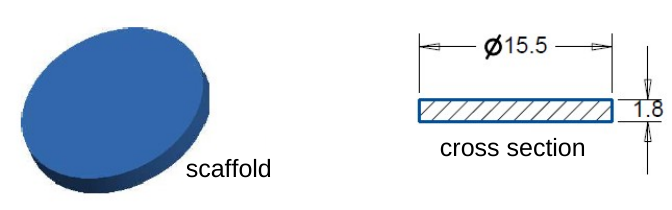}
		\caption{Scaffold dimensions (lengths in [mm])}
	\end{subfigure}
	\caption{Sketch of in-vitro experiment}\label{fig:experiment}
\end{figure}

This section is concerned with the detailed derivation of a new mathematical model for 
the major biological processes that characterize the cell colonization in a scaffold for meniscus
regeneration. We start by addressing the connection between the micro-, meso- and macroscopic scales
and then extend the model by mechanical effects that stem from the surrounding perfusion chamber. Fig.~\ref{fig:experiment} shows the experimental setup.

\subsection{A multiscale approach connecting subcellular and population levels}\label{sec:multiscale}

We consider a first model for the dynamics of human mesenchymal stem cells (hMSCs) differentiating into and interacting with chondrocytes, under chemical, topological, and mechanical environmental influences. Chondrocytes produce ECM and hyaluron, also uptaking the latter. The cells migrate, differentiate, and proliferate inside an artificial scaffold with given topology, which does not infer resorbtion by the cells and whose fibers are impregnated with hyaluron, that acts as a (nondiffusing) chemoattractant for the hMSCs. The differentiation of hMSCs to chondrocytes and the phenotype preservation of the latter are induced and sustained by a differentiation medium. We start at the microscopic level of single cells and therewith associated receptor binding dynamics, then pass through the mesoscale of cell distribution functions, and obtain by a parabolic upscaling effective equations for the dynamics of macroscopic population densities. The method follows closely that in \cite{conte2023mathematical,conte2021mathematical,corbin2021modeling}.\\[-2ex]

\noindent
{\bf Microscopic, subcellular scale.} We account for binding of hMSC receptors to their ligands available in the extracellular space. Here we focus on interactions with the chemoattractant (hyaluron) and with ECM and provide on this level less details for the exchange of chondrocytes with their environment. \\[-2ex]

\noindent
Let $y_1$ denote the amount of hMSC receptors occupied with hyaluron (of density $h$ and reference density $H$) and $y_2$ be the amount of hMSC receptors bound to ECM (of density $\tau$ and reference density $K$). Simple receptor binding kinetics then takes the form
\begin{align*}
&(R_0-y_{1}-y_{2}) +\frac{h}{H} \underset{k_{1}^{-}}{\stackrel{k_{1}^{+}}{\rightleftharpoons }}y_1\\
&(R_0-y_{1}-y_{2}) +\frac{\tau}{K} \underset{k_{2}^{-}}{\stackrel{k_{2}^{+}}{\rightleftharpoons }}y_2,
\end{align*}
where $R_0$ represents the total amount of bound receptors on an hMSC membrane; we assume $R_0$ to be constant. Then we get the ODE system
\begin{align*}
&\dot y_1=k_1^+\frac{h}{H}(R_0-y_1-y_2)-k_1^-y_1\\
&\dot y_2=k_2^+\frac{\tau}{K}(R_0-y_1-y_2)-k_2^-y_2,
\end{align*}
with $k_j^+$ an $k_j^-$ denoting attachment and respectively detachment rates of hMSC to hyaluron ($j=1$) and ECM ($j=2$). For simplicity we assume $k_1^-=k_2^-=:k^-$. Let us denote $y:=y_1+y_2$. Then we get 
\begin{equation}\label{ODE-y}
\dot y=\left (k_1^+\frac{h}{H}+k_2^+\frac{\tau}{K}\right )(R_0-y)-k^-y
\end{equation}
Rescaling $y/R_0\leadsto y\in (0,1)$ further simplifies the notation. Since receptor binding is very fast compared to the overall dynamics of cell migration and proliferation, we assume it to quickly reach the equilibrium and only deal with the steady-state of the above equation:
\begin{equation*}
y^{*} = \frac{k_{1}^{+} \frac{h}{H} +k_{2}^{+} \frac{\tau}{K}}{k_{1}^{+} \frac{h}{H} +k_{2}^{+} \frac{\tau}{K}+ k^-}.
\end{equation*}
We denote by $z:=y^*-y$ the (very small) deviation of $y$ from $y^*$ and proceeed as in \cite{conte2023mathematical,conte2021mathematical,corbin2021modeling,engwer2015glioma,engwer2016effective} to obtain from \eqref{ODE-y} 
\begin{equation*}
\dot{z}= -zB(h,\tau) +\frac{k^-}{(B(h,\tau))^2} v\cdot \nabla _xB(h,\tau):=G(z,h,\tau),
\end{equation*}
with $B(h,\tau):=k_{1}^{+} \frac{h}{H}+k_{2}^{+} \frac{\tau}{K}+k^-$.\\[-2ex]

\noindent
\textbf{Mesoscopic level.} The dynamics of cell density distributions for hMSCs and chondrocytes is described by way of kinetic transport equations. Let $p_1(t,x,v,z)$ denote the density of hMSCs sharing at time $t>0$ and position $x\in \R^n$ the velocity regime $v\in V_1=s_1\Sbb$ and the deviation $z\in Z= (y^*-1,y^*)$ from the equilibrium receptor binding state. Likewise, $p_2(t,x,v)$ represents the density of chondrocytes with velocity $v\in V_2 =s_2 \Sbb$. The positive constants $s_1$ and $s_2$ are the speeds of hMSCs and chondrocytes, respectively. The KTEs for $p_1$ and $p_2$ then write
\begin{align}
&\partial_tp_1+\nabla _x\cdot (vp_1)+\partial_z(G(z,h,\tau)p_1)=\mathcal L_1[\lambda_1 (z)]p_1+\text{(de)differentiation \& proliferation}\label{KTE:ADSCs}          \\
&\partial_tp_2+\nabla _x\cdot (v p_2)=\mathcal L_2[\lambda_2]p_2+\text{(de)differentiation}.\label{KTE-chondro}
\end{align}
The first terms on the right hand sides of \eqref{KTE:ADSCs} and \eqref{KTE-chondro} characterize the reorientation of hMSCs and chondrocytes, respectively. Concretely, choose for the turning operators
\begin{align}
&\mathcal{L} _1[\lambda_1(z)]p_1(t,x,v,z):=- \lambda_1(z)p_1(t,x,v,z)+\lambda_1(z) \int_{V_1} K_1(x,v)p(t,x,v',z) dv'\label{turnop-1}\\
&\mathcal{L} _2[\lambda _2]p_2(t,x,v):=-\lambda_2p_2(t,x,v)+\lambda_2\int_{V_2}K_2(x,v)p_2(t,x,v') dv',\label{turnop-2}
\end{align}
with the turning rates $\lambda_1(z) = \lambda_{10} - \lambda_{11} z \geq 0$ as in \cite{conte2023mathematical,conte2021mathematical,engwer2015glioma,engwer2016effective}, where $\lambda_{10},\lambda_{11}>0$ are constants and $\lambda_2>0$ is a constant, too. For the turning kernels we take into account (as in \cite{conte2023mathematical,conte2021mathematical,engwer2015glioma,engwer2016effective,painter2013mathematical}) the anisotropy of the scaffold fibers and choose $K_j(x,v):=q(x,\hat{v})/\omega_j$ ($j=1,2$), where $\hat{v} = \frac{v}{|v|}$ and $q(x, \theta)$ with $\theta \in \mathbb{S}^{n-1}$ is the orientational distribution of the scaffold fibers, normalised by $\omega _j= s_j^{n-1}$. We assume the tissue to be undirected, hence $q(x, \theta)= q(x, -\theta)$ for all $x$ in $\mathbb{R}^n$ and $\theta \in \Sbb $ . We introduce the notations
\begin{align*}
\mathbb{E}_q(x)&= \int_ {\mathbb{S}^{n-1} }\theta q(x,\theta )d\theta\qquad (\text{observe that }\mathbb{E}_q =0)\\
\mathbb{V}_q(x)&= \int_ {\mathbb{S}^{n-1} }(\theta-\mathbb{E}_q)\otimes (\theta-\mathbb{E}_q)  q(x,\theta ) d\theta.
\end{align*}
The turning operators in \eqref{turnop-1}, \eqref{turnop-2} take the form
\begin{align*}
\mathcal{L} _j[\lambda_j]p_j:=\lambda_j\left (\frac{q(x,\hat{v})}{\omega_j}\int_{V_j} p_jdv-p_j(v)\right )\qquad (j=1,2).
\end{align*}
We denote by $c_1(t,x):=\iint _{V_1\times Z}p_1(t,x,v,z)d(v,z)$ and $c_2(t,x):=\int _{V_2}p_2(t,x,v)dv$ the macroscopic densities of hMSCs and chondrocytes, respectively.\\[-2ex] 

\noindent
The source terms on the right hand sides in \eqref{KTE:ADSCs} and \eqref{KTE-chondro} describe proliferation and growth/decay due to differentiation of hMSCs into chondrocytes and dedifferentiation of the latter. Such processes need a substantially longer time to happen when compared to cell migration, therefore will require a rescaling by a sufficiently small factor $\eps^2$. We assume that the cell phenotype is neither determined by the cell direction nor by the orientation of fibers in the scaffold, but only depends on macroscopic quantities in the extracellular space. Then \eqref{KTE:ADSCs} and \eqref{KTE-chondro} take the form
\begin{align}
&\partial_tp_1+\nabla _x\cdot (vp_1)+\partial_z(G(z,h,\tau)p_1)=\mathcal L_1[\lambda_1 (z)]p_1+\eps^2\left(-\alpha_1(\chi,S)p_1+\alpha_2(\chi,S)p_2+\beta p_1(1-\frac{c_1}{C_1^*}-\frac{c_2}{C_2^*}) \right ) \label{KTE:ADSCs-full}          \\
&\partial_tp_2+\nabla _x\cdot (v p_2)=\mathcal L_2[\lambda_2]p_2+\eps^2\left(\alpha_1( \chi,S)p_1-\alpha_2(\chi,S)p_2\right ),\label{KTE-chondro-full}
\end{align}
where $C_j^*$ denotes the carrying capacity of the cell population $j$ ($j=1,2$), $\beta >0$ is the constant growth rate of hMSCs, and $\alpha _1( \chi,S),\alpha _2(\chi,S)$ represent the rates of differentiation of hMSCs into chondrocytes and dedifferentiation of the latter, respectively. The functions $\alpha_j$ depend on the concentration $\chi$ of the  differentiation medium. Moreover, mechanical and chemical effects influence cell  de(differentiation) \cite{aufderheide2004mechanical,fahy2018mechanical,janmey2007cell,lee2006influence} and can be included upon letting $\alpha_j$ depend on concentrations of such chemicals and a stress-related quantity $S$ \cite{Prendergast97}, e.g. $S=\frac{\sigma}{\mu_1}+\frac{\vartheta }{\mu_2}$, with $\sigma $ and $\vartheta$ denoting known maximum shear stress and interstitial fluid speed, respectively, and $\mu_j>0$ ($j=1,2$) being constants \cite{andreykiv2008simulation,huiskes1997biomechanical}.\\[-2ex]

\noindent
The high dimensionality of this system would require expensive numerical simulations, therefore we deduce in the next subsection a system of macroscopic equations for the dynamics of the two cell populations. An effective macroscopic system is also more conducive for assessing the relevant behavior of hMSCs and chondrocytes in interaction with hyaluron and (newly produced) ECM. \\[-2ex]

\noindent
Let us introduce the following moments of $p_1$ with respect to the 'activity' variable $z$ and velocity $v$:
\begin{align*}
m(t,x,v):= \int_Z p_1 (t,x,v,z)dz, \qquad 	m^{z}(t,x,v):= \int_Z zp_1 (t,x,v,z)dz, \qquad M^{z}(t,x):=	 \int_{V_1} m^{z}(t,x,v)dv.
\end{align*}
Due to the fact that $z$ is very small we will neglect moments of $p_1$ w.r.t. $z$ for higher orders. This will ensure the subsequent informal moment closure.\\[-2ex]

\noindent
\textbf{Upscaling and macroscopic level.} We perform a parabolic scaling of the time and space variables: $t\leadsto \eps^2t$ and $x\leadsto \eps x$. Applying this to \eqref{KTE:ADSCs-full}, \eqref{KTE-chondro-full} leads to
\begin{align}
\eps ^2\partial _tp_1+\eps \nabla_x\cdot (vp_1)&+\partial_z\left (\left (-zB(h,\tau)+\eps \frac{k^-}{(B(h,\tau))^2} v\cdot \nabla _xB(h,\tau)\right )p_1\right )\notag \\ 
&=\mathcal L_1[\lambda_1 (z)]p_1+\eps^2\left(-\alpha_1({ \chi},S)p_1+\alpha_2({ \chi},S)p_2+\beta p_1(1-\frac{c_1}{C_1^*}-\frac{c_2}{C_2^*}) \right )\label{KTE:ADSCs-full-rescaled}\\
\eps ^2\partial _tp_2+\eps \nabla_x\cdot (vp_2)&=\mathcal L_2[\lambda_2]p_2+\eps^2\left(\alpha_1({\chi},S)p_1-\alpha_2({ \chi},S)p_2\right ).\label{KTE-chondro-full-rescaled}
\end{align}
We assume $p_1$ to be compactly supported in the phase space $\R^n\times V_1\times Z$. Integrating \eqref{KTE:ADSCs-full-rescaled} w.r.t. $z$ gives
\begin{align}
\eps ^2\partial_tm+\eps \nabla_x\cdot (vm)=-\lambda_{10}\left (m-\frac{q}{\omega _1}c_1\right )&+\lambda _{11}\left (m^z-\frac{q}{\omega _1}M^z\right )\notag \\
&+\eps^2\left (-\alpha_1({ \chi},S)m+\alpha_2({ \chi},S)p_2+\beta m(1-\frac{c_1}{C_1^*}-\frac{c_2}{C_2^*})\right ).\label{eq10}
\end{align}
Now multiply \eqref{KTE:ADSCs-full-rescaled} by $z$ and integrate w.r.t. $z$ to get
\begin{align}
\eps ^2\partial_tm^z+\eps \nabla_x\cdot (vm^z)+m^zB(h,\tau)-&\eps \frac{k^-}{(B(h,\tau))^2} v\cdot \nabla _xB(h,\tau)m=-\lambda_{10}\left (m^z-\frac{q}{\omega _1}M^z\right )-\eps^2\alpha_1({ \chi},S)m^z\notag \\
&{ +\eps^2\alpha_2({ \chi},S)p_2(y^*-2)}+\eps^2\beta m^z\left(1-\frac{c_1}{C_1^*}-\frac{c_2}{C_2^*}\right ). \label{eq11}
\end{align}
We perform Hilbert expansions of the $p_1$-moments and of $p_2$ and $c_2$:
\begin{align*}
m= \sum_{j=0}^{\infty}\eps^j m_j, \quad m^{z}= \sum_{j=0}^{\infty}\eps^j m^{z}_j,\quad c_1=	\sum_{j=0}^{\infty}\eps^j c_{1j},\quad M^{z}=	\sum_{j=0}^{\infty}\eps^j M^{z}_j,\quad 
p_2=	 \sum_{j=0}^{\infty}\eps^j p_{2j},\quad
c_2=	\sum_{j=0}^{\infty}\eps^j c_{2j}.
\end{align*}
Equating powers of $\eps$ in \eqref{KTE-chondro-full-rescaled}, \eqref{eq10}, \eqref{eq11} we get\\[-2ex]

\noindent
$\eps ^0$:
\begin{align}
&0=\lambda _2(\frac{q}{\omega _2}c_{20}-p_{20})\label{four}\\
&0=-\lambda_{10}\left (m_0-\frac{q}{\omega _1}c_{10}\right )+\lambda_{11}\left (m_0^z-\frac{q}{\omega _1}M_0^z\right)\label{fuenf}\\
&m_0^zB(h,\tau)=-\lambda_{10}\left (m_0^z-\frac{q}{\omega _1}M_0^z\right )\label{seven}
\end{align}
$\eps^1$:
\begin{align}
&\nabla_x\cdot (vp_{20})=\lambda_2\left (\frac{q}{\omega _2}c_{21}-p_{21}\right)\notag \\
&\nabla_x\cdot (vm_0)=-\lambda_{10}\left (m_1-\frac{q}{\omega _1}c_{11}\right )+\lambda_{11}\left (m_1^z-\frac{q}{\omega _1}M_1^z\right)\label{eps1-beflast} \\
&\nabla_x\cdot (vm_0^z)+m_1^zB(h,\tau)-\frac{k^-}{(B(h,\tau))^2} v\cdot \nabla _xB(h,\tau)m_0=-\lambda_{10}\left (m_1^z-\frac{q}{\omega _1}M_1^z\right )\label{eps1-last}
\end{align}
$\eps^2$ (from \eqref{KTE-chondro-full-rescaled}, \eqref{eq10}):
\begin{align*}
\partial_tp_{20}+\nabla_x\cdot (vp_{21})=&\lambda_2\left (\frac{q}{\omega _2}c_{22}-p_{22}\right)+\alpha_1({ \chi},S)p_{10}-\alpha_2({ \chi},S)p_{20}\\
\partial_tm_0+\nabla_x\cdot (vm_1)=&-\lambda_{10}\left (m_2-\frac{q}{\omega _1}c_{12}\right )+\lambda_{11}\left (m_2^z-\frac{q}{\omega _1}M_2^z\right)-\alpha_1({ \chi},S)m_0+\alpha_2({ \chi},S)p_{20}\\
&+\beta m_0\left(1-\frac{c_{10}}{C_1^*}-\frac{c_{20}}{C_2^*}\right ).
\end{align*}
Integrate \eqref{seven} w.r.t. $v$ to obtain $M_0^z=0$. Substitute this in \eqref{seven} to follow $m_0^z=0$. From \eqref{fuenf} and \eqref{four} follows 
\begin{equation}
m_0=\frac{q}{\omega _1}c_{10},\qquad p_{20}=\frac{q}{\omega _2}c_{20}.
\end{equation}
Integrating \eqref{eps1-last} w.r.t. $v$ leads to $M_1^z=0$. Then from \eqref{eps1-last} we obtain
\begin{equation}\label{eq:(18)}
m^{z}_1 =  \frac{k^-}{B(h,\tau)^2(B(h,\tau)+\lambda_{10})}v\cdot \nabla_xB(h,\tau)\frac{q}{\omega_1}c_{10}.
\end{equation}
Now apply the operator $\mathcal{L}_1[\lambda_{10}]$ to $m_1(t,x,v)=\int _Zp_{11}(t,x,v,z)dz$:
\begin{align*}
\mathcal{L}_1[\lambda_{10}]m_1&=\lambda_{10}\left (\frac{q}{\omega _1}c_{11}-m_1\right )\\
&=\nabla_x\cdot(vm_0)-\lambda _{11}m_1^z\qquad (\text{due to \eqref{eps1-beflast}}).
\end{align*}
The compact Hilbert-Schmidt operator on the weighted space $L^2_{\frac{q}{\omega _1}}(V_1):=\{w\in L^2(V_1)\ :\ \int _{V_1}w^2\frac{dv}{\frac{q}{\omega _1}}<\infty \}$ has kernel $\left <\frac{q}{\omega _1}\right >:=\text{span}(\frac{q}{\omega _1})$, thus its pseudo-inverse can be determined on $\left <\frac{q}{\omega _1}\right >^\perp$. We obtain (for more details refer e.g., to \cite{OthmerHillen2000,engwer2015glioma}):
\begin{align}\label{eq-m1}
m_1=-\frac{1}{\lambda_{10}}\left (\nabla_x\cdot(vm_0)-\lambda _{11}m_1^z\right ),\quad \text{thus}\quad c_{11}=0.
\end{align}
Analogously,
\begin{align}\label{eq-p21}
p_{21}=-\frac{1}{\lambda_{2}}\left (\nabla_x\cdot(vp_{20})\right ),\quad \text{thus}\quad c_{21}=0.
\end{align}
Now integrate the $\eps^2$-equations w.r.t. $v$ to obtain
\begin{align}
&\partial_tc_{10}+\nabla_x\cdot \int _{V_1}vm_1\ dv=-\alpha_1({ \chi},S)c_{10}+\alpha_2({ \chi},S)\frac{\omega_1}{\omega_2}c_{20}+\beta c_{10}\left(1-\frac{c_{10}}{C_1^*}-\frac{c_{20}}{C_2^*}\right )\label{eq-c10}\\
&\partial_tc_{20}+\nabla_x\cdot \int _{V_2}vp_{21}\ dv=\alpha_1({ \chi},S)\frac{\omega_2}{\omega_1}c_{10}-\alpha_2({ \chi},S)c_{20}.\label{eq-c20}
\end{align}
In virtue of \eqref{eq-m1} we evaluate
\begin{align*}
&\nabla_x\cdot \int _{V_1}vm_1=-\nabla\nabla :\left (\mathbb D_1c_{10}\right )+\nabla \cdot \left (\frac{k^-\lambda _{11}}{B(h,\tau)^2(B(h,\tau)+\lambda_{10})}\mathbb D_1\nabla B(h,\tau)c_{10}\right )\\
&\nabla_x\cdot \int _{V_2}vp_{21}=-\nabla\nabla :\left (\mathbb D_2c_{20}\right ),
\end{align*}
with 
\begin{align}
&\mathbb D_1(x)=\frac{1}{\lambda_{10}}\int _{V_1}v\otimes v\frac{q(x,\hat v)}{\omega _1}dv=\frac{s_1^2}{\lambda_{10}}\int _{\Sbb }\theta \otimes \theta q(x,\theta ),\label{diff-tensor-1}\\
&\mathbb D_2(x)=\frac{1}{\lambda_{2}}\int _{V_2}v\otimes v\frac{q(x,\hat v)}{\omega _2}dv={ \frac{\lambda_{10}}{\lambda_2}\left (\frac{s_2}{s_1}\right )^2\mathbb D_1(x)}.\label{diff-tensor-2}
\end{align}
\noindent
Plugging these into \eqref{eq-c10}, \eqref{eq-c20} and neglecting higher order terms in the Hilbert expansions of $c_1$ and $c_2$ (see also \eqref{eq-m1}, \eqref{eq-p21}) we obtain the macroscopic reaction-diffusion-taxis equations (RDTEs)
\begin{align}
\partial_tc_{1}-\nabla\nabla :\left (\mathbb D_1c_{1}\right )+\nabla \cdot \left (\frac{k^-\lambda _{11}}{B(h,\tau)^2(B(h,\tau)+\lambda_{10})}\mathbb D_1\nabla B(h,\tau)c_{1}\right )&=-\alpha_1({ \chi},S)c_{1}+\alpha_2({ \chi},S)\frac{\omega_1}{\omega_2}c_{2}\notag \\
&\quad +\beta c_{1}\left(1-\frac{c_{1}}{C_1^*}-\frac{c_{2}}{C_2^*}\right )\label{macro-c1}\\
\partial_tc_{2}-\nabla\nabla :\left (\mathbb D_2c_{2}\right )&=\alpha_1({ \chi},S)\frac{\omega_2}{\omega_1}c_{1}-\alpha_2({ \chi},S)c_{2}.\label{macro-c2}
\end{align}
Thereby, $\nabla\nabla :(\mathbb D c)=\nabla \cdot (\nabla \cdot \mathbb Dc+\mathbb D\nabla c)$ represents myopic diffusion; the Fickian diffusion with tensor $\mathbb D\in \R^{n\times n}$ infers a drift correction with convection velocity $\nabla \cdot \mathbb D$.\\[-2ex]

\noindent
We supplement the above RDTEs with the dynamics of { differentiation medium with concentration $\chi $ 
\begin{equation}\label{macro-chi}
\partial _t\chi=D_\chi\Delta \chi -a_\chi (c_1+c_2)\chi 
\end{equation}
representing diffusion and uptake by both cell phenotypes} along with those for hyaluron concentration $h$ and ECM density $\tau$, none of which is supposed to diffuse:
\begin{align}
\partial_th& = { -\gamma_1hc_1}-\gamma_2hc_2+\frac{c_2}{1+c_2}\label{macro-h}  \\
\partial_t\tau& = -\delta c_1\tau +c_2,\label{macro-k}
\end{align}
with the source terms on the right hand side characterizing production, degradation, and uptake of $h$ and $\tau$ due to $c_1$ and $c_2$.\\[-2ex]

\noindent
Thus far we dealt with $x\in \R^n$; however, we should actually consider a bounded region $\Omega_p \subset \R^n$ in 
which cells, ECM, and hyaluron are evolving.\footnote{Subsequently $\Omega _p$ will be a bounded and convex subdomain of a domain $\Omega \subset \R^n$ with sufficiently regular boundary $\partial \Omega$.} This raises the need for boundary conditions, which can be obtained as e.g. in \cite{plaza}, see also \cite{corbin2021modeling,conte2023mathematical} for similar deductions. Hence, we supplement system \eqref{macro-c1}-\eqref{macro-k} with no-flux boundary conditions
\begin{align}
&\left (\mathbb D_1\nabla c_1+\left(\nabla \cdot \mathbb D_1-\frac{k^-\lambda _{11}}{B(h,\tau)^2(B(h,\tau)+\lambda_{10})}\mathbb D_1\nabla B(h,\tau)\right )c_{1}\right )\cdot \nu =0\quad \text{on}\quad \partial \Omega_p\label{macro-c1-BC}\\
&\left (\nabla \cdot \mathbb D_2c_2+\mathbb D_2\nabla c_2\right )\cdot \nu =0\quad \text{on}\quad \partial \Omega_p.\label{macro-c2-BC}
\end{align}
{ We also consider a no-flux boundary condition for $\chi$:
\begin{equation}\label{BC-chi}
\nabla \chi \cdot \nu =0\quad \text{ on }\partial \Om_p.
\end{equation}
} 

\noindent
{ For the initial conditions we consider $h$ to be uniformly distributed in space: $h_0(x)=\tilde h_0\frac{1}{|\Omega_p|}\mathbbm{1}_{\Omega _p}(x)$, with $\tilde h_0$ a given constant. The initial densities $c_{1,0}$ and $c_{2,0}$ of hMSCs and chondrocytes, respectively, are considered to be Gaussians, as the cells are placed at the center of the upper scaffold interface, from where they spread. We take $\tau (0,x)=0$, as there is initially no newly formed ECM. The differentiation medium is provided at several different times during the experiment, each time the same overall quantity, which is supposed to quickly diffuse within the whole domain $\Omega _p$, thus we consider it to be uniformly distributed: $\chi(t\in T_\chi,x)=\chi_0\frac{1}{|\Omega_p|}\mathbbm{1}_{\Omega _p}(x)$, where $T\chi=\{0,2,3,6,9,12,15,18,21,24\}$ is the set containing the time points (in days) where the available differentiation medium is substituted by a new one, with concentration $\chi_0$.}

\subsection{Including mechanical effects}\label{subsec:mechanics}

The cells migrate, proliferate, and (de)differentiate within an articial scaffold integrated in a 3D printed perfusion chamber which is embedded in a bioreactor\footnote{Experimental setup designed and implemented at DITF Denkendorf by M. Dauner, M. Doser, C. Linti, and A. Ott}. As mentioned in above, mechanical stress is an important factor leading to cell (de)differentiation.  To account for it, the bioreactor is endowed with an alternating fluid flowing through the perfusion chamber tubes and releasing the pressure.\\[-2ex] 

\noindent
Thus, the previously obtained macroscopic model for the dynamics of cells, hyaluron, and newly produced ECM is to be supplemented with fluid dynamics and therewith induced deformations of the scaffold. These, in turn, will affect the evolution of $c_1$, $c_2$, $h$, and $\tau $, the coupling being realized by way of the (de)differentiation rates $\alpha _1$ and $\alpha_2$ which appear on the right hand sides of \eqref{macro-c1} and \eqref{macro-c2}. We proceed as in \cite{Grosjean23} and consider the tissue (ECM and scaffold) as a poroelastic medium which we model by the Biot equations (see e.g. \cite{Ambarts})
\begin{subequations}
	\begin{align}
	\rho_s \pt_{tt} \boldsymbol{\eta}_p - \nabla\cdot {\boldsymbol{\sigma}_p}(\boldsymbol{\eta}_p,p_p) & =  0, 
	\label{eq:B61} \\
	\partial_t \left( \frac{1}{M} p_p + \nabla \cdot (\alpha \boldsymbol{\eta}_p) \right)
	+ \nabla \cdot \mathbf{u}_p & =  0,  \label{eq:B62}
	\end{align}
\end{subequations}
with the displacement field $\boldsymbol{\eta}_p(t,x)$ and the pressure $p_p(t,x)$ in a domain
$\Omega_p \subset \mathbb{R}^n$, with additional boundary and initial conditions.
Here, $\rho_s$ stands for the solid phase density, while $M$ and $\alpha$ are Biot's modulus and coefficient, respectively. The stress tensor $\boldsymbol{\sigma}_p$ is given by 
\begin{equation}
\label{eq:sigmap}\boldsymbol{\sigma}_p(\etaeta_p,p_p)=\boldsymbol{\sigma}_e(\etaeta_p) - \alpha p_p I,  \quad  \mathrm{and }  \quad \boldsymbol{\sigma}_e(\etaeta_p)=\lambda_p (\nabla \cdot \etaeta_p)\ I + 2 \mu_p D(\etaeta_p),
\end{equation}
with $\lambda_p$ and $\mu_p$ being the Lam\'e parameters, whereas the fluid flux satisfies Darcy's law 
\begin{equation}
\label{eq:darcyeq}
\mathbf{u}_p = -\mathbb{K}(\nabla p - \rho_f \mathbf{g})/\mu,
\end{equation} with permeability matrix $\mathbb{K}$, fluid phase density $\rho_f$, and  viscosity $\mu$. Additionally, a Stokes flow is considered:
\begin{equation}
\label{eq:unstokes}
\rho_f \pt_t \uu_f  -  \nabla \cdot \boldsymbol{\sigma}_f(\uu_f,p_f)=0,   \quad \mathrm{ and }  \quad  \nabla \cdot \uu_f=0, \quad \mathrm{ in }\  \Omega_f,
\end{equation}
for the fluid stress tensor given by $\boldsymbol{\sigma}_f(\uu_f,p_f):=-p_f I + 2 \mu D(\uu_f),  \textrm
{with } D(\uu_f)=\frac{1}{2} (\nabla \uu_f + \nabla \uu_f^T)$ and where  $\rho_f$ stands for the fluid phase density. \\[-2ex]

\noindent
The cells and the deformation of tissue evolve in $\Omega _p$, while $\Omega_f$ is the domain occupied by the fluid. It holds that $\Om=\Om_p\cup \Om_f$ and we denote by  $\mathbf{n}_f$ the outward unit normal vector to the boundaries $\Gamma_f=\Gamma_I \cup \Gamma_{f,W} \cup \Gamma_{in} \cup \Gamma_{out},$ where $\Gamma_I$ represents the interface between $\Omega_f$ and $\Omega_p$, $\Gamma_{f,W}$ represents the wall boundaries of $\Om_f$, and $\Gamma_{in}$, $\Gamma_{out}$ are the inflow and outflow boundaries, respectively. \\[-2ex]

\noindent
The boundary conditions for the fluid are
\begin{align*}
&\uu_f = 0 \textrm{ on } \Gamma_{f,W},\\
&\boldsymbol{\sigma}_f \ \nn_f = -p_{in}(t) \ \nn_f \textrm{ and } \uu_f \times \nn_f = 0\textrm{ on } \Gamma_{in},\\
&\boldsymbol{\sigma}_f \ \nn_f =0 \textrm{ on } \Gamma_{out}.
\end{align*}
The pressure boundary condition at the inflow is set to 
$ p_{in}(t)=~p_{max}  \sin(\pi t)$. \\[-2ex]

\noindent
Likewise, we denote $\Gamma_p:=\Gamma_{p,W} \cup \Gamma_I$, where $\Gamma_{p,W}$ represents the 'wall boundary' of the region $\Om_p$. We require 
\begin{align*}
p_p=0 \text{ and }\boldsymbol{\eta}_p=0\textrm{ on }  \Gamma_{p,W}.
\end{align*}
At the interface $\Gamma_I$ we prescribe the following conditions:
	\begin{itemize}
		\item mass conservation: \begin{equation}
		\label{massconv}
		\uu_f \cdot \nn_f + ( \pt_t \etaeta_p + \uu_p ) \cdot \nn_p =0.
		\end{equation}
		\item balance of stresses:\begin{align}
		\label{stressbal}
		&- (\boldsymbol{\sigma}_f \ \nn_f)\cdot \nn_f=(p_f - 2\mu D(\uu_f )) \nn_f ) \cdot \nn_f = p_p, \notag \\
		&    \boldsymbol{\sigma}_f \cdot \nn_f + \boldsymbol{\sigma}_p \cdot \nn_p= (2 \mu_f D(\uu_f) - p_f I )\cdot \nn_f + (\lambda_p \nabla \cdot \etaeta + 2\mu_p D(\etaeta) - \alpha p_p) \cdot \nn_p = 0.
		\end{align}
		\item the Beavers-Joseph-Saffman (BJS)
		condition modeling slip with friction:
		\begin{equation}
		\label{BJSeq}
		-(\boldsymbol{\sigma}_f \nn_f)_T = \underbrace{\mu_f \alpha \mathbb{K}^{-1/2}}_{\mathbf{\alpha}_{BJS}} (\uu_f - \pt_t \etaeta_p)_T,
		\end{equation}
		where $\mathbf{\alpha}_{BJS}$ is the slip rate coefficient and $(\cdot)_T$ is the tangential component.
	\end{itemize}

\begin{figure}[h!]\begin{center}
\includegraphics[width=5cm]{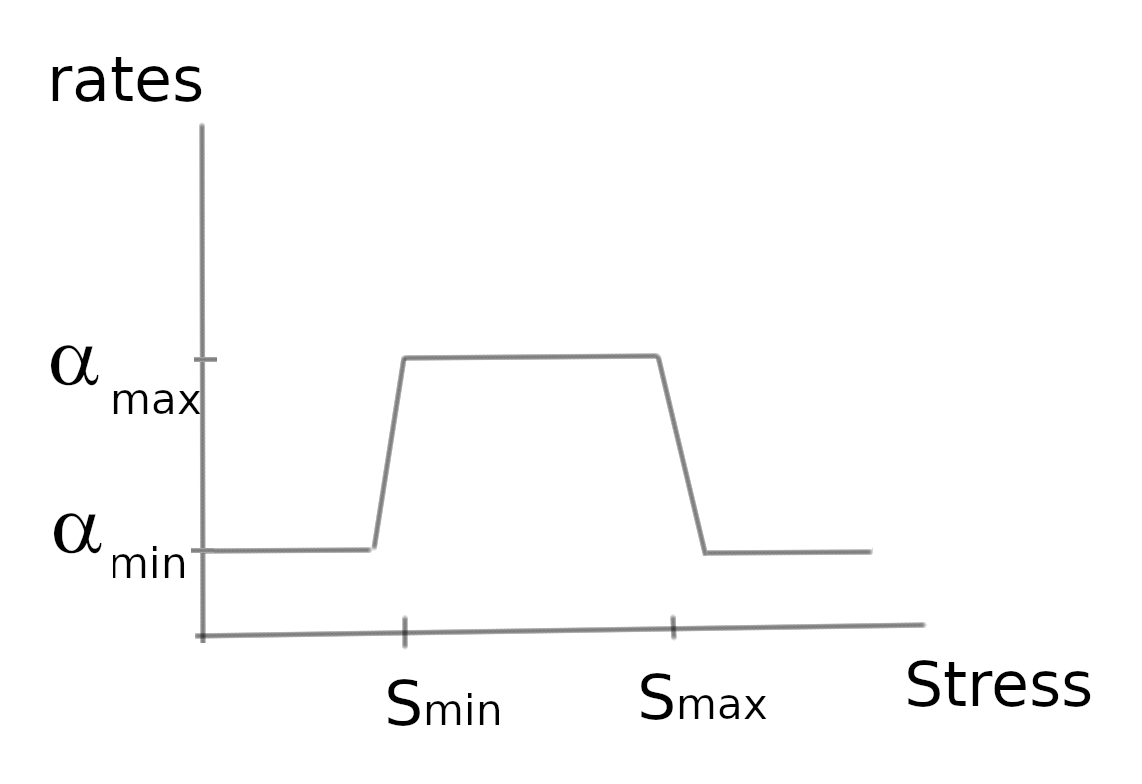}
\caption{Shape of mapping connecting mechanical stimulus $S$ with (de)differentiation rates}
\label{fig:S-and-alphas}\end{center}	
\end{figure}

\noindent
For the coupling via the rates $\alpha_1$ and $\alpha_2$ we propose an approach inspired by \cite{andreykiv2008simulation,Prendergast97}, where the mechanical stimulus $S$ is defined as a linear combination between a strain-related quantity and the fluid velocity. The authors showed that $S$ would favor chondrocyte differentiation and ECM synthesis if it got between two values, $S$min and $S$max. We thus propose a mapping
between $\boldsymbol{\sigma}_p$ and $\alpha_1,\ \alpha_2$, which thus become dependent on time, space, and on the stress. Such dependency is shown in Figure \ref{fig:S-and-alphas}.

\section{Experimental data}\label{sec:experiments} 
This section summarizes the experimental work that has been carried out so far, starting with
the fabrication of nonwoven scaffolds and then proceeding with image analysis and a 
biomechanical characterization.

\subsection{Scaffold fabrication}

\noindent
A PET (polyethylene terephthalate) multifilament was extruded and stretched, the resulting  diameter of the individual filament was $17 \mu m$. The filaments were cut to fibers of about $60 mm$ length and the fibers were thermally treated in a drying oven to avoid shrinkage in the subsequent process steps. A non-woven fiber web was formed from these fibers with a carding machine by winding a defined number of pile layers up to the desired weight. The fiber web was then mechanically solidified by a needle machine. In order to achieve a predetermined porosity, the resulting needle felt was passed through a defined gap between two heated calender rolls and the required fleece thickness was thermally set. Thus, non-woven scaffolds with a thickness of about $1.8 mm$ and a slightly different porosity were generated: one with a weight of $267-292 g/m^2$ (LW, $88\%$ porosity) and one with $400-420 g/m^2$ (HW, $85\%$ porosity).

\subsection{Imaging analysis of simple scaffolds}


\noindent
Six scaffold samples were scanned by micro computed tomography using a Skyscan1172 device. A source voltage of 40 kV and a current of 250\,\textmu A were used. The angular step width in the rotation was $0.3$ degrees resulting in 1202 projection images. The voxel spacing is 3.98\,\textmu m. \\[-2ex]

\begin{figure*}
	\begin{subfigure}{\textwidth}
		\centering
		\begin{tikzpicture}[
		spy using outlines={
			rectangle,
			magnification=5,
			width=7cm,
			height=8.15cm,
			connect spies}]
		\node[inner sep=0pt] {\includegraphics[width=.5\textwidth]{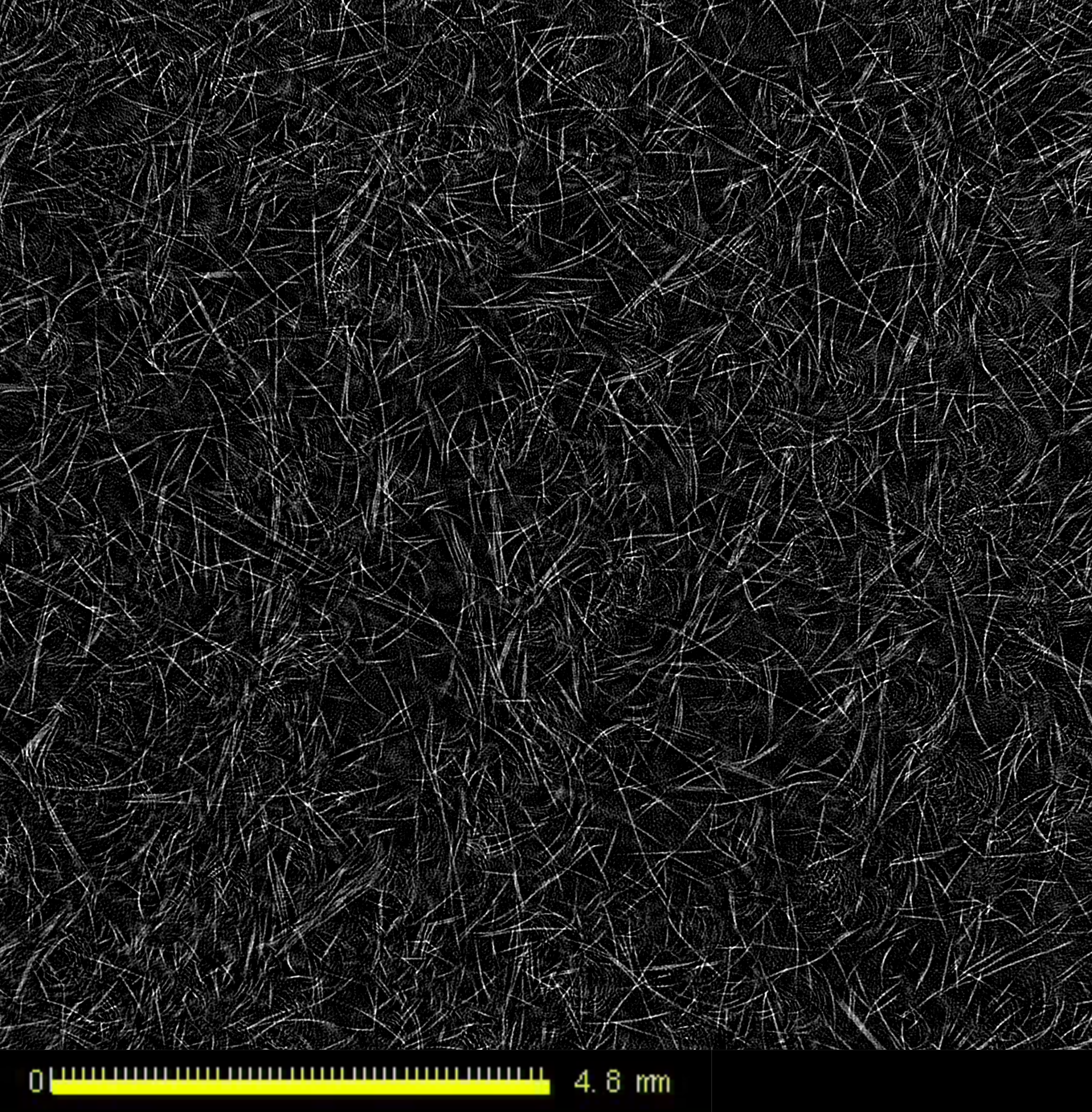}};
		\spy[red] on (-1.5cm,0cm) in node at (.5\textwidth,0cm);
		\end{tikzpicture}
	\end{subfigure}
	\caption{Sectional image of the CT scan of sample 1 (left). The area marked by a red box is magnified 5 times (right) to enhance the visibility of the fiber structure.}
	\label{Fig:CT_Section}
\end{figure*}

\noindent
We cropped the images such that the cuboid field of view is entirely inside the cylindrical specimen (approx. 2600\,$\times$\,2600\,$\times$\,300\,voxels). A sectional image is shown in Figure \ref{Fig:CT_Section}. In a preprocessing step, we smoothed the CT image with a Gaussian filter (filter mask size: 7\,$\times$\,7\,$\times$\,7, $\sigma = 6.0$, reflective edge treatment) to remove noise while preserving edges. These parameters were chosen according to the fiber diameter which corresponds to approximately $r \approx 6$ voxels.\\[-2ex]



\noindent
To achieve a coarse segmentation of the fiber system, we deployed Otsu's thresholding method~\cite{otsu79} with a factor of $1.25$. For further refinement, we applied a median filter (filter mask size: 3\,$\times$\,3\,$\times$\,3, reflective edge treatment), then a morphological closing operation (structuring element size: 5\,$\times$\,5\,$\times$\,5), and, subsequently, the median filter again. Lastly, we removed connected components of sizes smaller than $50$\,voxels because these are likely too small to constitute single fibers.\\[-2ex]

\noindent
For each fiber voxel, we estimated the fiber orientation with the Hessian matrix of the  grey value image~\cite{eberly94, ohser-schladitz09book}. The approach is based on the idea that the fiber direction is aligned with the direction of minimal curvature. Therefore, the eigenvector belonging to the smallest eigenvalue of the Hessian is used as the fiber direction estimate. For computing the Hessian, we used $\sigma = 3$, which equals the fibre radius, since this choice has been shown to be optimal in~\cite{wirjadi16, pinter18}.\\[-2ex]

\noindent
In the next step, the fiber direction in the typical fiber point (in Palm sense, see \cite{KMS95}) is modelled by an angular central Gaussian (ACG) distribution. The density of this distribution model on the unit sphere in $\mathbb{R}^3$ is given by
\begin{equation}\label{ACG-distribution}
f_A(\theta )=  \frac{1}{4 \pi { |\det A|^{\frac{1}{2} }}} (\theta^TA^{-1}\theta )^{-\frac{3}{2}}, \quad \theta\in \mathbb S^2.
\end{equation}


\noindent
The ACG distribution can be interpreted as the distribution of the direction of a $\mathcal{N}_3(0, A)$-distributed random vector or, alternatively, as the distribution
of a random point uniformly distributed on the ellipsoid ${\theta  : \theta^T A^{-1}\theta = 1}$.
A sample for estimating the parameter matrix $A$ is obtained by the fiber directions of a subsample of foreground (fiber) voxels in the image. Subsampling reduces the amount of data to an acceptable level.
Moreover, neighboring voxels are highly dependent as they often belong to the same fiber. By subsamplig, we were able to create a nearly independent data set, which is a prerequisite for the parameter estimation. Voxels were sampled by applying independent Bernoulli trials with $p=1/100\,000$ to all voxels. Prior to the estimation, we rotated the orientations in each sample such that its principal direction was aligned with the x-axis. This way, we ensured comparability between samples. \\[-2ex]

\begin{figure*}
	\begin{subfigure}{\textwidth}
		\centering
		\includegraphics[width=0.45\textwidth]{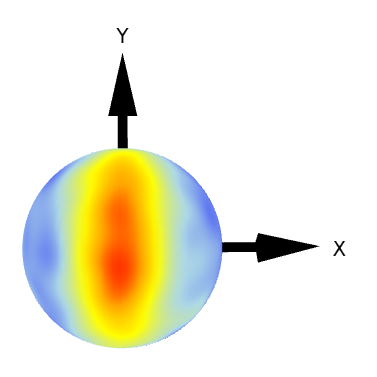}
		\includegraphics[width=0.45\textwidth]{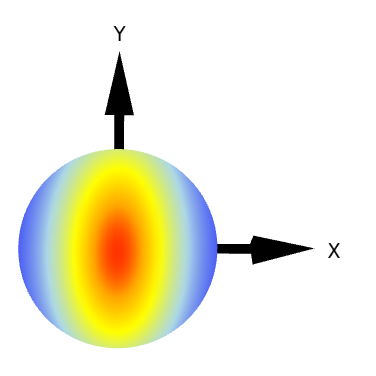}
	\end{subfigure}
	\caption{Spherical density plots of the empirical fiber orientation distribution (left) and of the fitted ACG model (right) for sample 1.}
	\label{ACG-plot}
\end{figure*}

\noindent
We calculated the maximum likelihood estimate for the ACG parameter matrix $A$ using the fixed-point algorithm proposed by Tyler~\cite{tyler1987StatisticalAnalysisAngular}. For the results, see Table~\ref{tab:param} and Fig.~\ref{ACG-plot} for a visualization. To validate the goodness of fit, we inspected QQ plots for the longitude and colatitude~\cite{fisher1987StatisticalAnalysisSpherical}. The model quantiles were obtained by the empirical quantiles of a simulated sample of the fitted ACG model~\cite{Rfast}. 
For all samples, the QQ plots indicate a good fit, see Fig.~\ref{qq-first}~-~\ref{qq-last}.\\[-2ex]

\begin{table}[!t]
	\begin{center}
		\begin{tabular}{|l||c|c|c|c|c|c|}
			\hline
			&$A_{11}$& $A_{12}$& $A_{13}$ &  $A_{22}$& $A_{23}$ &  $A_{33}$\\
			\hline
			\hline
			Sample 0 & \phantom{-}1.697 & \phantom{-}0.023 & -0.028 & \phantom{-}0.873 & -0.031 & \phantom{-}0.324\\
			\hline
			Sample 1 & \phantom{-}1.687 & \phantom{-}0.003 & \phantom{-}0.064 & \phantom{-}0.914 & -0.007& \phantom{-}0.399 \\
			\hline
			Sample 2 & \phantom{-}1.483 & \phantom{-}0.018 & \phantom{-}0.023 & \phantom{-}1.077 & \phantom{-}0.022& \phantom{-}0.440 \\
			\hline
			Sample 3 & \phantom{-}1.341 & -0.005 & \phantom{-}0.158 & \phantom{-}1.083 & \phantom{-}0.036 & \phantom{-}0.576 \\
			\hline
			Sample 4 & \phantom{-}1.698 & \phantom{-}0.018 & \phantom{-}0.016 & \phantom{-}0.911 & -0.001 & \phantom{-}0.391 \\
			\hline
			Sample 5 & \phantom{-}1.493 & -0.006 & -0.010 & \phantom{-}1.097 & -0.011 & \phantom{-}0.410 \\
			\hline
			Sample 6 & \phantom{-}1.479 & -0.010 & \phantom{-}0.051 & \phantom{-}1.067 & -0.021 & \phantom{-}0.454\\
			\hline
		\end{tabular}
		\caption{ACG parameter estimates for the fiber orientation of the scaffolds. \label{tab:param}}
	\end{center}
\end{table}

\noindent
To determine the cell diffusion tensors \eqref{diff-tensor-1}, \eqref{diff-tensor-2} we need to assess the mesoscopic orientation distribution of fibers $q(x,\theta)$ from the scaffold data. The analysis of the CT images does not give evidence of a strong local variation of the fiber direction distribution. Therefore, we will use $q(x, \theta) = q(\theta) = f_A(\theta)$ as determined in \eqref{ACG-distribution}. The second moments can be determined according to the method in \cite{ospald}: with $q(x,\theta)=f_A(\theta)$ we compute 
\begin{equation}\label{elliptic-integr-D}
\mathbb D_\beta=\int _{\mathbb S^2}\theta \otimes \theta q(x,\theta)d\theta =c_{A,\beta}\int \limits _0^\infty \prod \limits _{i=1}^3(b_i+\zeta )^{-\frac{\beta _i+1}{2}}d\zeta ,
\end{equation}
where $\beta =(\beta _1,\beta_2,\beta _3)^T$ is a multiindex with $\sum\limits _{i=1}^3\beta _i=2$ used to specify the entries in $\mathbb D_\beta$, $b_i$ ($i=1,2,3$) are the entries in $A^{-1}=\text{diag}(b_1,b_2,b_3)$, and 
\begin{equation*}
c_{A,\beta}:=\frac{|\det A|^{-1/2}}{4}\prod \limits_{i=1}^3\frac{\beta_i!}{(\beta_i/2)!}.
\end{equation*}
The entries of $\mathbb D_\beta$ can thus be obtained by computing the elliptic integrals in \eqref{elliptic-integr-D} above. For instance, $\mathbb D_{(2,0,0)}=\mathbb D_{11}=\frac{|\det A|^{-1/2}}{2}\int \limits _0^\infty(b_1+\zeta )^{-3/2}(b_2+\zeta)^{-1/2}(b_3+\zeta)^{-1/2}d\zeta $.

\subsection{Biomechanical characterization of PET scaffolds}

\subsubsection*{Materials and Methods} 
\paragraph*{Scanning electron microscopy analysis of three-dimensional PET scaffolds}
Three-dimensional (3D) PET scaffolds were cut (diameter: $4 mm$, thickness: $1.85 mm$), process for critical point-drying. Afterwards, the samples fixed with carbon tape, and coated with a $20 nm$ gold-palladium film to become electrically conductive. High-resolution images were acquired from PET surfaces in an environmental scanning electron microscope (SEM, S-5200, Hitachi, Tokyo, Japan) with an acceleration voltage of $15 kV$ for qualitative assessment of the distribution of the fibres.

\paragraph*{Indentation mapping}
A biomechanical indentation mapping of the PET scaffolds was performed to investigate the homogeneity of the scaffold's material properties. PET scaffolds with two different grammage ranges: $267-292 g/m^2$ (LW) and $400-420 g/m^2$ (HW) were tested. PET scaffolds sterilized by a dose of 25 kGy of gamma irradiation, as recommended for terminal sterilization of medical products, were also tested. All samples were analysed twice; before and after a 60 min hydration period with 10 ml phosphate buffer saline (PBS). Briefly, a multiaxial mechanical tester (MACH-1 v500css, Biomomentum Inc., Laval, QC, Canada) equipped with a 17N load cell was used to perform spatial stress relaxation tests on the surface of the PET scaffolds in accordance to an established algorithm \cite{ref:ulm2}. First, the build- in camera-registration system was used to define a pattern with measurement points on the surface (n=6 measurement points). On each measurement point a non-destructive indentation relaxation test was performed using a spherical indenter ($\emptyset=2mm$, indentation depth: $15\% h0$ (initial sample height), velocity: $\sim 5\% h0/s$, relaxation time: $10s$). At each measurement point, the structural stiffness of the PET scaffolds was calculated \cite{ref:ulm3}.

\begin{figure}[hbt]
\centering
\includegraphics[width=0.65\textwidth]{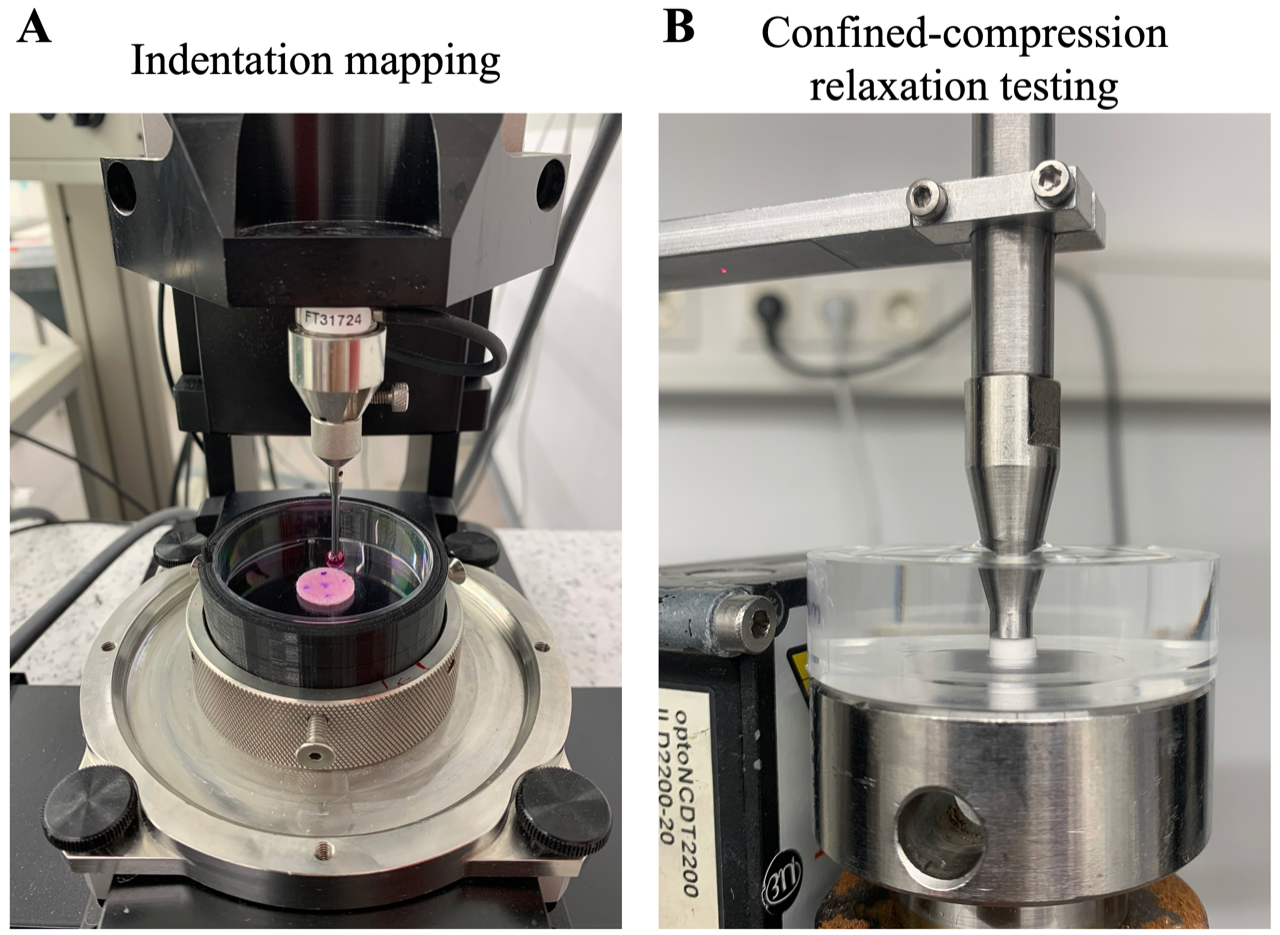}
\caption{The PET scaffolds were mechanically investigated using two different testing methods: A) spatial indentation mapping was performed on the surface of the entire PET scaffold. Subsequently, cylindrical samples were punched out of the scaffolds to further analyse the biomechanical properties under B) confined-compression conditions.}
\label{fig:ulm1}
\end{figure}

\paragraph*{Multi-step confined compression relaxation test}
Following the spatial indentation mapping \cite{ref:ulm1}, the viscoelastic properties of the PET scaffolds were further investigated under confined conditions in a material testing machine (Z10, ZwickRoell, Germany). 4,5 Cylindrical samples ($\emptyset = 4.6 mm$) were punched out at the before tested measurement points using a biopsy punch (Stiefel Laboratories Inc, UK). For the stress-relaxation test, the sample was placed in a custom-made confined compression testing chamber filled with PBS \ref{fig:ulm1}. One side of the PET scaffolds was facing the impermeable bottom of a measuring chamber, and the opposite side was facing a porous ceramic (Al2O3) cylinder to allow free fluid flow. The measuring chamber had the same diameter ($\emptyset = 4.6 mm$) as the samples, thus, confining their radial deformation. The materials testing machine equipped with a stainless-steel punch induced an initial preload of $0.1 N$ to ensure the same testing conditions at the beginning of each test, while the thickness of the samples (h0) was automatically registered. Then, the samples were loaded to three consecutive, incremental strain levels ($\eps_i$) of $0.1$, $0.15$, and $0.2$ at an individual sample loading rate of $100\%h0/min$. Each strain level was held constant for 15 minutes to ensure an equilibrium state was reached. Sample strain was continuously measured and controlled using a laser displacement transducer (optoNCDT 2200-20, Micro-Epsilon GmbH \& Co. KG, Germany, $0.3\mu m$ resolution, $\pm 0.03\%$ accuracy). The resulting force was measured by a $20 N$ load cell (ZwickRoell, Germany) \cite{ref:ulm4,ref:ulm5}.\\[-2ex] 

\noindent
Data analysis was performed using a custom-made MATLAB R2020a (MathWorks Inc, USA) script. At each strain rate, the equilibrium's modulus $E_{eq}$ representing the matrix stiffness was calculated by the quotient of the stress at equilibrium ($\sigma _{t\to \infty}$) and the applied strain ($\eps _i$): 

\begin{equation}\label{eq1:ulm}
E_{eq}=\frac{\sigma _{t\to \infty}}{\eps _i},\qquad \eps_i=0.1, 0.15, 0.2.
\end{equation}
\noindent
Solving the diffusion equation of Mow et al. by means of nonlinear least squares regression strain rates yields the hydraulic permeability $k$ as a measure of the resistance to fluid flow through the PET scaffolds for all three strain levels. The aggregate modulus $H$ indicates a combined modulus of both the fluid and the solid phase of the PET scaffolds \cite{ref:ulm6}: 

\begin{equation}\label{eq2:ulm}
\sigma_t=\sigma _{t\to \infty}+2H\eps_ie^{-\left (\frac{\pi}{h_0}\right )^2Hkt}
\end{equation}

\paragraph*{Statistical analysis} 
GraphPad Prism 9 software (GraphPad Software, Inc, La Jolla, CA, USA) was used for the
statistical analysis. Normal distribution was assessed with the Shapiro-Wilk test. The data of
the normal indentation tests were normally distributed. For the comparison between dry and
hydrated conditions paired t-test was performed, while for the comparison between the non-
sterile and sterile PET scaffolds and the comparison of the LW and HW PET samples unpaired
t-test with Welch's correction was used. The results of the confined tests were analysed using
Mann-Whitney testing. The level of significance was defined as $p < 0.05$.

\subsubsection*{Results}

\begin{figure}[hbt]
	\centering
	\includegraphics[width=0.65\textwidth]{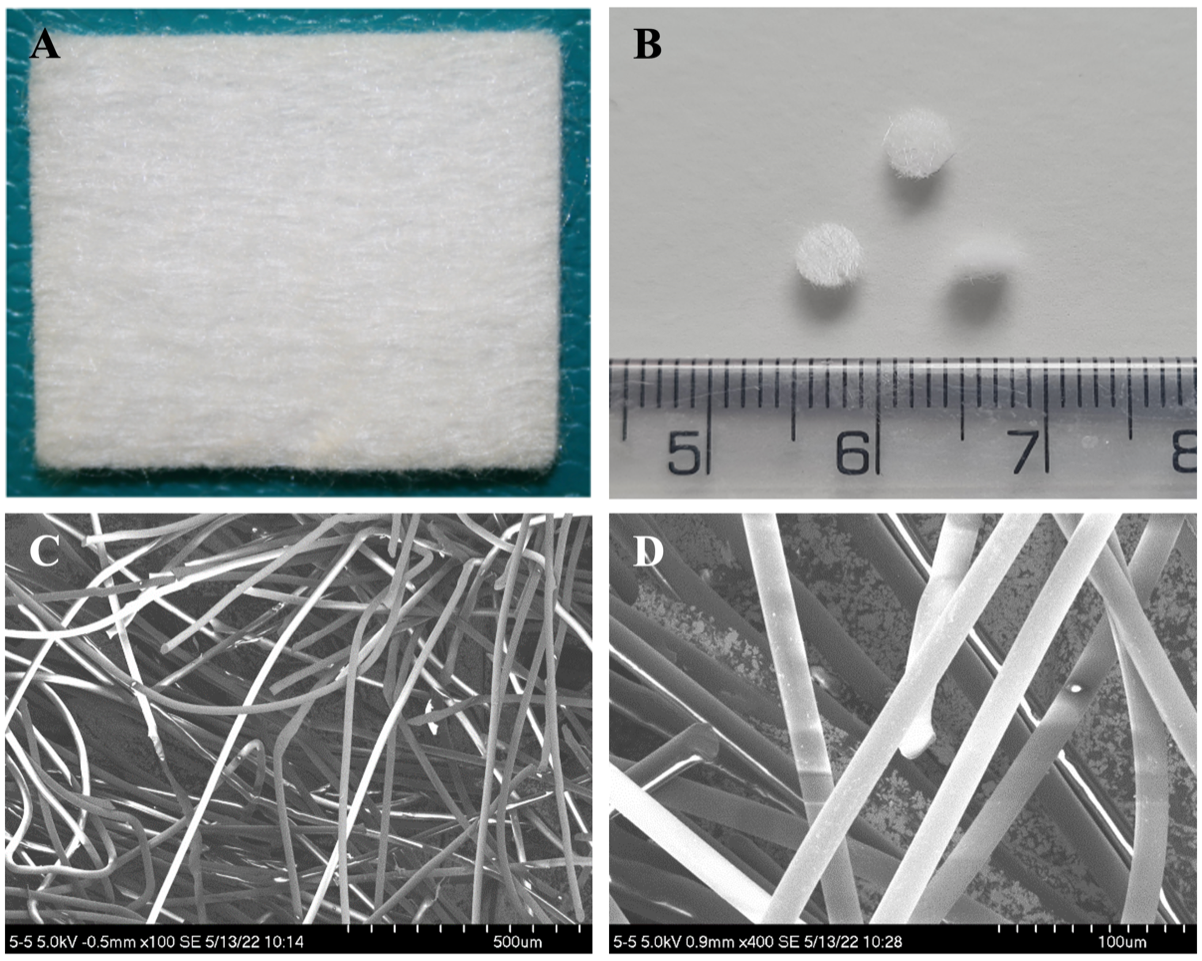}
	\caption{3D PET scaffolds. A) Top view of PET sheet and B) top and peripheral view of PET
		punches. C, D) Representative scanning electron microscopy images from interior cross-section of PET samples at two different magnifications (scale bars indicate $500 \mu m$ and $100 \mu m$)}
	\label{fig:ulm2}
\end{figure}

\paragraph*{PET biomechanical characterization} 
The scaffold punches collected from the PET (Figure \ref{fig:ulm2} A, B) were prone to loosing  fibres with
handling and displayed unorganized fibre distribution under SEM imaging (Figure \ref{fig:ulm2} C, D).

\begin{figure}[hbt]
	\centering
	\includegraphics[width=0.85\textwidth]{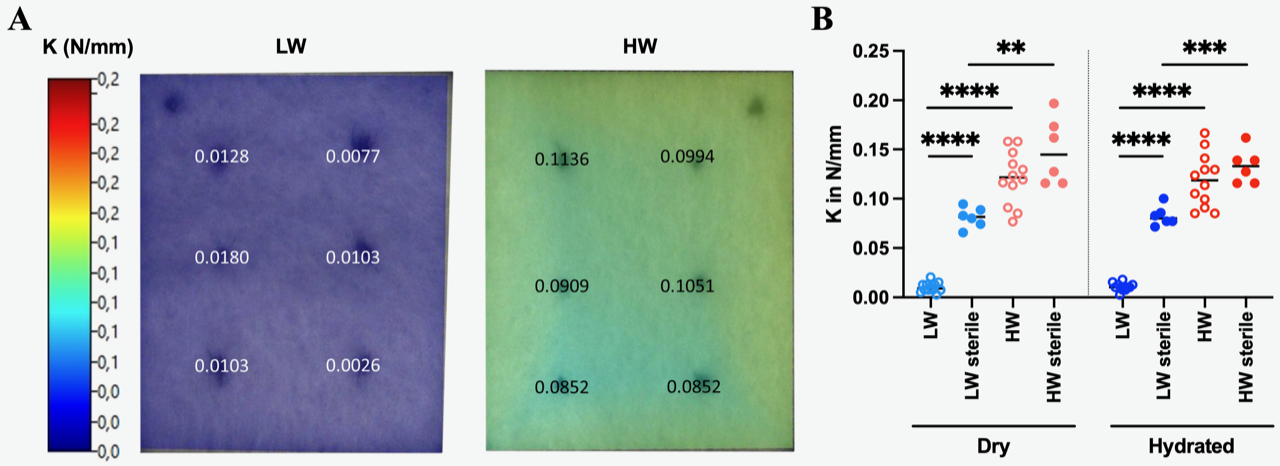}
	\caption{Results of the indentation mapping. A) Representative stiffness mappings of a LW non-sterile and HW non-sterile PET scaffold. B) Structural stiffness ($K$ in $N/mm$) are shown for the low weight (LW) and heigh weight (HW) PET scaffolds in dry and hydrated condition, each for the non-sterile and sterile samples. $n = 6$ (sterile), $n =12$ (non-sterile). $**\ p < 0.01$, $***\ p < 0.001$, $****\ p < 0.0001$.}
	\label{fig:ulm3}
\end{figure}

\noindent
No differences in the structural stiffness between the dry and hydrated state were observed for
the PET scaffolds (Figure \ref{fig:ulm3}). When comparing the non-sterile and sterile LW samples,
significantly higher K values were found for the sterile PET scaffolds both under dry and hydrated
condition ($p<0.0001$), whereas no differences were found between non-sterile and sterile HW
scaffolds. The pairwise comparisons of the LW and HW scaffolds revealed significantly higher $K$
values for the HW PET scaffolds both in the non-sterile and sterile groups (dry: $p<0.01$; hydrated:
$p<0.001$).

\begin{figure}[hbt]
	\centering
	\includegraphics[width=0.85\textwidth]{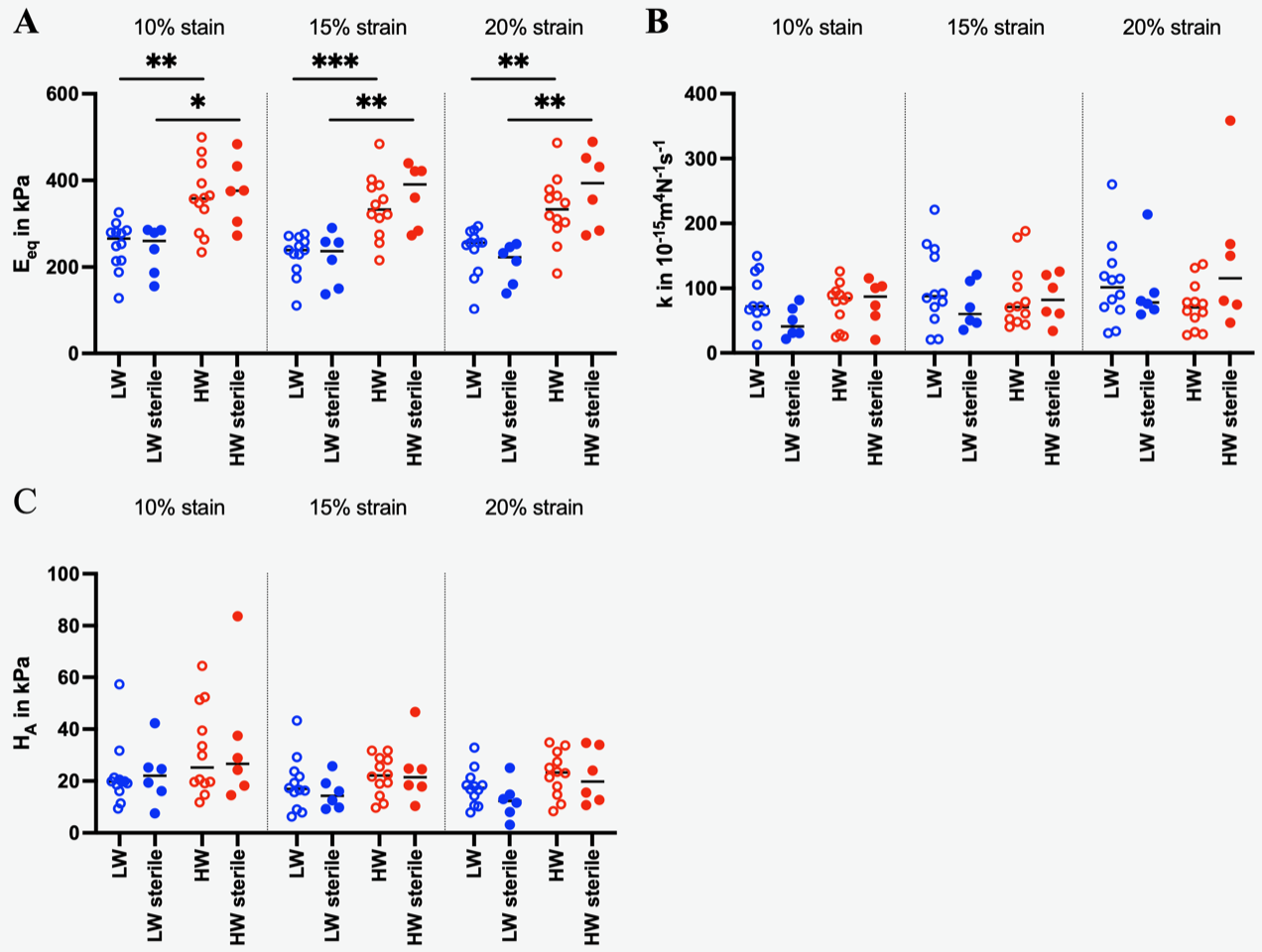}
	\caption{Results of the confined compression tests at $10\%$, $15\%$ and $20\%$ strain for the low weight (LW) and high weight (HW) PET scaffolds under non-sterile and sterile conditions. A) Equilibrium modulus ($E_{eq}$ in $kPa$), B)
		permeability ($k$ in $10^{-15}m^4N^{-1}s^{-1}$), and C) aggregate modulus ($H_A$ in $kPa$). $n = 6$ (sterile), $n = 12$ (non-sterile). $*\ p <0.05$, $**\ p < 0.01$, $***\ p < 0.001$.}
	\label{fig:ulm4}
\end{figure}

\noindent
No differences in the equilibrium modulus ($E_{eq}$) were found between the non-sterile and sterile
samples in both the LW and HW PET scaffolds (Figure \ref{fig:ulm4} A). At all strain levels, the non-sterile
and sterile LW PET scaffolds indicated significantly lower $E_{eq}$ values compared to the respect
HW scaffolds ($p<0.05$). For the permeability ($k$), no differences were found between the non-
sterile and sterile samples at none of the three strain levels (Figure \ref{fig:ulm4} B). The comparison
between the LW and HW PET revealed no significant differences in all conditions. The
aggregate modulus $H_A$ was not significantly different when comparing the non-sterile and
sterile samples in none of the three strain levels, neither for the LW nor the HW scaffolds (Figure
\ref{fig:ulm4} C). No differences in $H_A$ were found when comparing the LW with the HW PET scaffolds in all three strain levels and in sterile versus non-sterile conditions.

\section{Numerical methods} \label{sec:numerics}

In this section, we summarize the numerical treatment of both the fluid flow simulation 
in the perfusion chamber and the cell dynamics in the scaffold. We start with 
the flow system \eqref{macro-c1}-\eqref{BC-chi}, supplemented with mechanical effects as described in Section \ref{subsec:mechanics}, and use FreeFem++ (version 4.1) for 
the spatial discretization. The codes are available on GitHub.\\[-2ex]

\noindent
Let us define the spaces 
\begin{equation}
\label{eq:spaces}
\begin{cases}
&V_f=\{\mathbf{v}_f \in H^1(\Omega_f)^d: \mathbf{v}_f=0 \textrm{ on } \Gamma_f\}, \\
&P_f=L^2(\Omega), \\
&V_p=H(div; \Omega_p),\\
&P_p= L^2(\Omega_p),\\
&\Xi_p= \{\boldsymbol{\xi}_p \in H^1(\Omega_p)^d: \ \boldsymbol{\xi}_p =0 \textrm{ on } \Gamma_p\}.
\end{cases}
\end{equation}
Using the variational formulation of \eqref{eq:B61}-\eqref{eq:B62}-\eqref{eq:darcyeq}- \eqref{eq:unstokes}, the following boundary terms at the interface $\Gamma_I$ 
between scaffold and surrounding fluid flow appear: 
\begin{equation}
\label{IGPF}
-(\boldsymbol{\sigma}_p \textbf{n}_p, \boldsymbol{\xi})_{\Gamma_I}+ (p_p\ \textbf{n}_p ,\mathbf{v}_p)_{\Gamma_I}-(\boldsymbol{\sigma}_f \textbf{n}_f, \mathbf{v}_f)_{\Gamma_I},
\end{equation}
with $\boldsymbol{\xi} \in \Xi_p$, $\mathbf{v}_p \in V_p$ and $\mathbf{v}_f \in V_f$.\\[-2ex] 

\noindent
We set all variables to $0$ for the initial conditions.
We consider a referential fixed domain such that the surface $\Gamma_I$ is perpendicular to the axis $(Oz)$, $\mathbf{n}_f=(0,0,-1)$ and the tangential vectors are given by $\mathbf{t}_1=(1,0,0)$ and $\mathbf{t}_2=(0,-1,0)$. Then, we apply Nitsche's method as in \cite{biotdarcystokes1}, which allows us to correctly impose the mass conservation \eqref{massconv} with penalization terms. We obtain the following variational problem: \\
Find $(\uu_f,p_f,\uu_p,p_p,\etaeta_p) \in (V_f,P_f,V_p,P_p,\Xi_p)$ such that for all  $(\mathbf{v}_f ,q_f,\mathbf{v}_p,q_p,\boldsymbol{\xi}_p) \in (V_f,P_f,V_p,P_p,\Xi_p)$
\begin{equation}
\label{eq:coupledweak}
\begin{cases}
(\rho_f \pt_t \uu_f, \mathbf{v}_f)_{\Omega_ f}  +  2 \mu_f (D(\uu_f ), D(\mathbf{v}_f ))_{\Omega_ f} - (p_f, \nabla \cdot \mathbf{v}_f)_{\Omega_f} + (q_f, \nabla \cdot \uu_f)_{\Omega_ f}  \\
+ (\rho_p \pt_{tt} \etaeta_p, \boldsymbol{\xi}_p)_{\Omega_ p} + (\lambda_p \nabla \cdot \etaeta_p,\nabla \cdot \boldsymbol{\xi}_p)_{\Omega_p}  + (2 \mu_p D(\etaeta_p), D(\boldsymbol{\xi}_p))_{\Omega_p} - (\alpha p_p , \nabla \cdot \boldsymbol{\xi}_p)_{\Omega_p} \\
+ (\mu K^{-1} \uu_p,\mathbf{v}_p)_{\Omega_p} - (p_p,\nabla \cdot \mathbf{v}_p)_{\Omega_p} \\
+ \pt_t (\frac{1}{M}p_p,q_p)_{\Omega_p} + (\alpha \pt_t \nabla \cdot \etaeta_p, q_p )_{\Omega_p} + (\nabla \cdot \uu_p, q_p)_{\Omega_p} \\
+ I_{\Gamma} = -(p_{in}(t) \nn_f ,\mathbf{v}_f)_{\Gamma_{in}}, \\
\end{cases}
\end{equation}
where 
\begin{align}
\label{endgamma}
I_{\Gamma}&=\int_{\Gamma_I} \nn_f \cdot \boldsymbol{\sigma}_f  \nn_f (\boldsymbol{\xi}_p+\mathbf{v}_p -\mathbf{v}_f)  \cdot \nn_f +\int_{\Gamma_I} \gamma \mu_f h^{-1} (\uu_f-\uu_p-\pt_t \etaeta_p) \cdot \nn_f (\mathbf{v}_f-\mathbf{\xi}_p-\mathbf{v}_p) \cdot \nn_f  \notag \\
&\quad  - \int_{\Gamma_I} \mathbf{\alpha}_{BJS} \textbf{t} \cdot (\uu_f-\pt_t \etaeta_p) (\boldsymbol{\xi}_p - \mathbf{v}_f) \cdot \textbf{t}
+ \int_{\Gamma_I} \nn_f \cdot (q_fI + 2 \mu D( \mathbf{v}_f))  \nn_f (\pt_t \etaeta_p + \uu_p -\uu_f)  \cdot \nn_f,
\end{align}
and where $h$ is the size of the mesh, and $\gamma$ a penalization. \\[-2ex]

\noindent
We proceed with direct simulations of the model for the cell dynamics \eqref{macro-c1}--\eqref{macro-k} on the scaffold, denoted $\Omega_p$. A numerical scheme should be locally mass conservative, thus we have decided to employ a first order Non-symmetric Interior Penalty discontinuous Galerkin (NIP dG) scheme in space  \cite{di2011mathematical}. 
We define a mesh $\mathcal{T}_h$ of $\Omega_p$ and seek solutions $c_1$, $c_2$ and $\chi$ in the broken polynomial space $\mathbb{P}^1_d( \mathcal{T}_h)$ given by $\mathbb{P}_d^1(\mathcal{T}_h):=\{ u \in L^2(\Omega_p)\ | \ \forall T \in \mathcal{T}_h, v_{|_T} \in \mathbb{P}_d^1(T) \}$, ~whereas we are looking for $h$ and $k$ on the classical $\mathbb{P}^1_d( \Omega_p)$ FE space.  
Multiplying by test functions $(\nu_{c1}, \nu_{c2}, \nu_{chi}, \nu_{h},\nu_{\tau})$ and integrating over $\Omega_p$, the system  \eqref{macro-c1}-- \eqref{macro-k}  becomes
\begin{equation}
\begin{cases}
\label{NIPsys1}
& (\pt_t c_1, \nu_{c1}) + ( \mathbb D_1 \nabla c_{1},\nabla \nu_{c1}) + ([c_1],\{\mathbb D_1 \nabla \nu_{c1}\})_{\Gamma} - ([\nu_{c1}],\{\mathbb D_1 \nabla c_1\})_{\Gamma} \\
& \quad \quad \quad  - (\textbf{v}\ c_1,\nabla \nu_{c1}) + ((\textbf{v} c_1)^{\uparrow},[\nu_{c1}])_{\pt \Omega_p}   \\
&   \quad\quad \quad  + (\alpha_1(\chi,S) c_1 -\frac{\omega_2}{\omega_1}\alpha_2(\chi,S) c_2  - \beta c_1(1-c_1-c_2), \nu_{c1}) + (\eta[c_{1}],[\nu_{c1}])_{\Gamma}=0,  \\
&(\pt_t c_2, \nu_{c2}) + ( \mathbb D_2 \nabla c_{2},\nabla \nu_{c2}) + ([c_2],\{\mathbb D_2 \nabla \nu_{c2} \})_{\Gamma}\\
& \quad \quad \quad - ([\nu_{c2}],\{ \mathbb D_2 \nabla c_2\})_{\Gamma} - (\frac{\omega_2}{\omega_1}\alpha_1(S) c_1 -\alpha_2(S) c_2, \nu_{c2}) +(\eta[c_{2}],[\nu_{c2}])_{\Gamma}= 0,  \\
&(\pt_t \chi, \nu_{\chi}) + ( D_{\chi} \nabla \chi,\nabla \nu_{\chi}) + ([\chi],\{D_{\chi} \nabla \nu_{\chi} \})_{\Gamma}- ([\nu_{\chi}],\{ D_{\chi} \nabla \chi\})_{\Gamma} \\ & \quad \quad \quad + (a_{\chi} (c_1+c_2)) +(\eta[\chi],[\nu_{\chi}])_{\Gamma}= 0,  \\
& (\pt_t h, \nu_h) + (\gamma_1\ h \ c_1,\nu_h) + ( \gamma_2 \ h \ c_2 ,\nu_h) - (\frac{c_2}{1+c_2},\nu_h)=0, \\
& (\pt_t \tau, \nu_{\tau}) + (\delta_1\ \tau \ c_1,\nu_{\tau}) - ( c_2,\nu_{\tau})=0, \\
& c_1(0) = c_1^0, \ c_2(0) = c_2^0, \ h(0) = h_0, \tau(0)=\tau_0.
\end{cases}
\end{equation}
Here, $\nabla$ refers to the broken gradient, $\Gamma $ represents all the interfaces of the mesh, $\eta$ is the penalization parameter, $\mathbf{v}=b_1 \nabla h+b_2 \nabla k$, $(\cdot, \cdot)$ refers to the $L^2(\Omega_p)$ inner product, $(\cdot)^{\uparrow}$ is the upwind flux, and $[\cdot]$ and $\{ \cdot \}$ refer to jumps and means. 
The nonlinear system \eqref{NIPsys1} has then been discretized in time by implicit Euler,
which needs Newton's method in each timestep. \\[-2ex]

\begin{figure}
	\begin{subfigure}{\textwidth}
		\centering
		\includegraphics[width=0.30\textwidth]{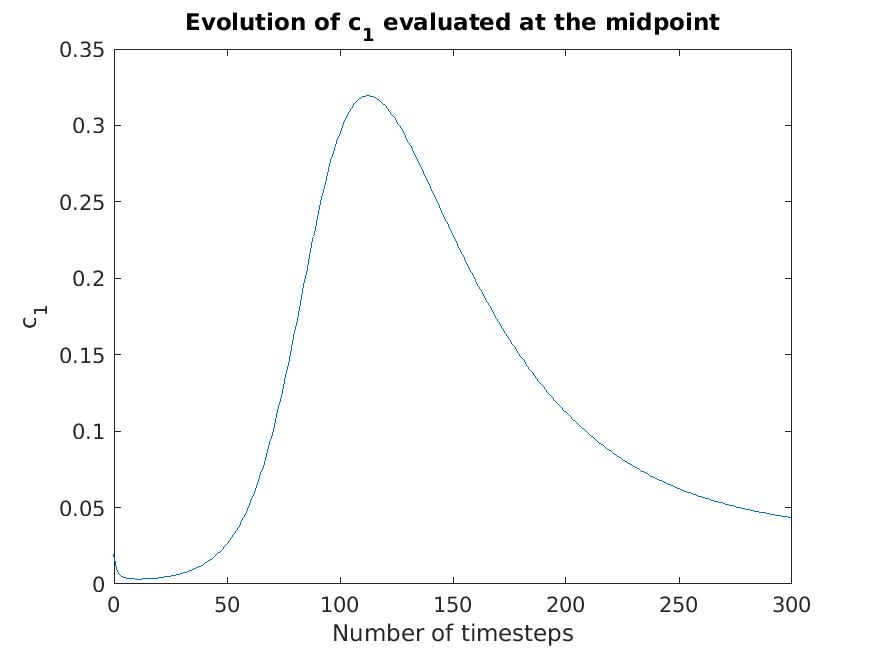}
		\includegraphics[width=0.30\textwidth]{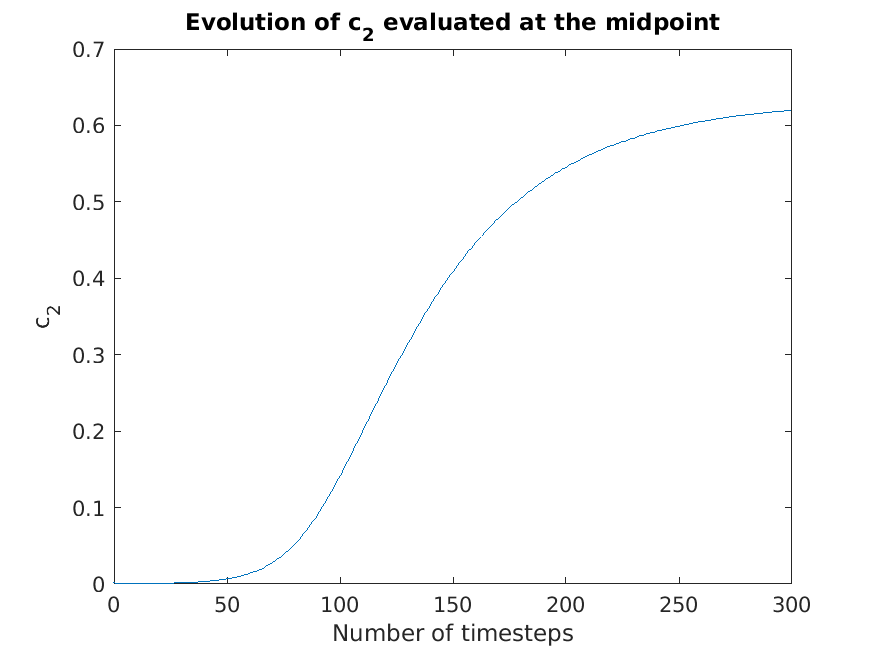}
		\includegraphics[width=0.30\textwidth]{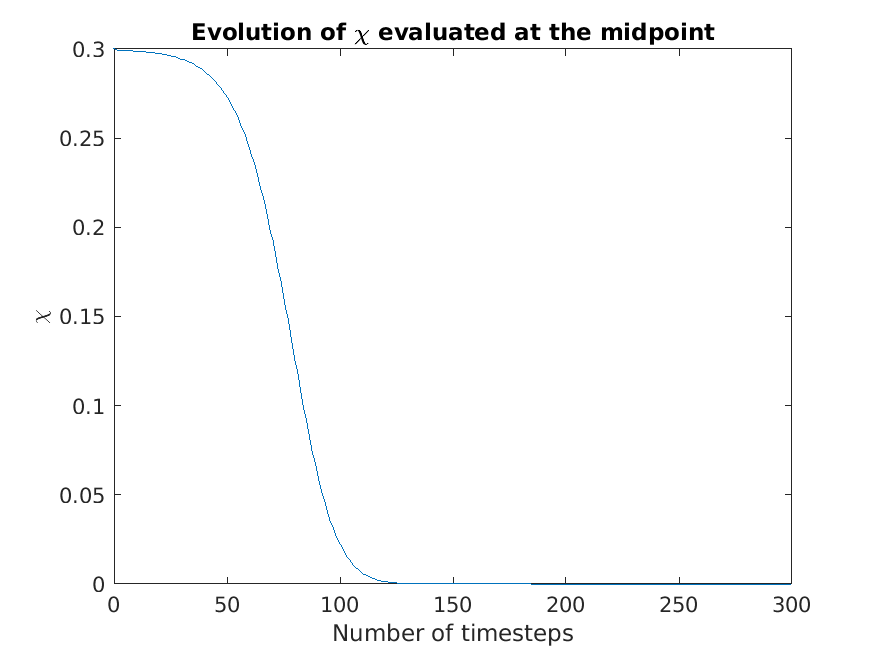}
		\includegraphics[width=0.30\textwidth]{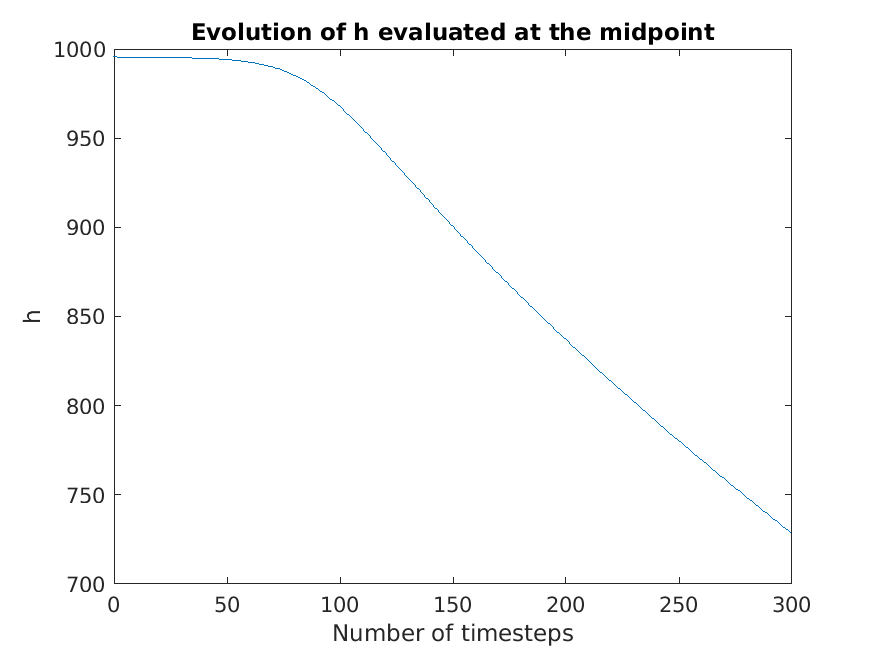}
		\includegraphics[width=0.30\textwidth]{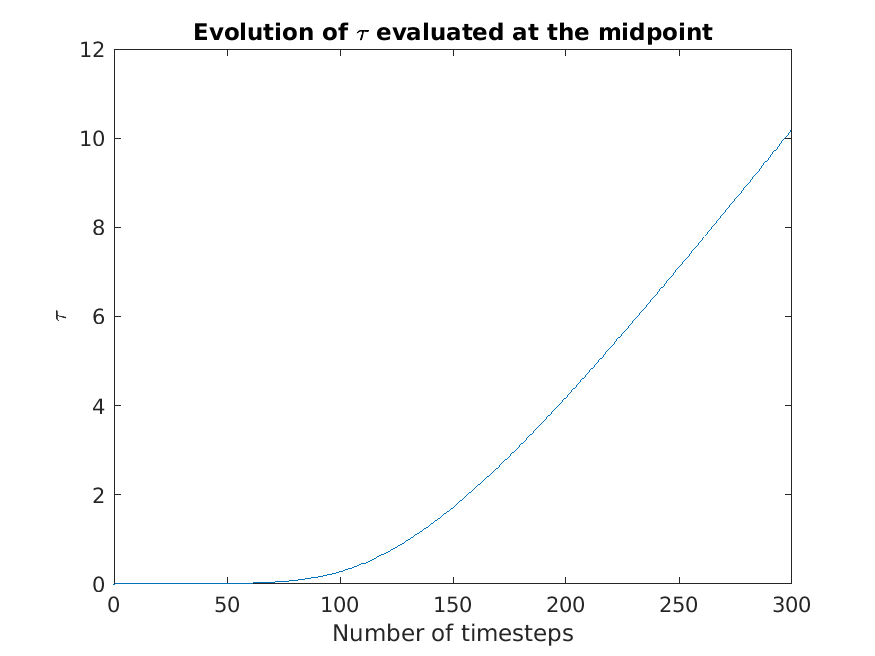}
		\caption{Evolution of $c_1$, $c_2$, $\chi$, $h$ and $\tau$ in the midpoint of the scaffold}
	\end{subfigure}
	\vspace*{4mm}
	
	\begin{subfigure}{\textwidth}
		\centering
		\includegraphics[width=0.47\textwidth]{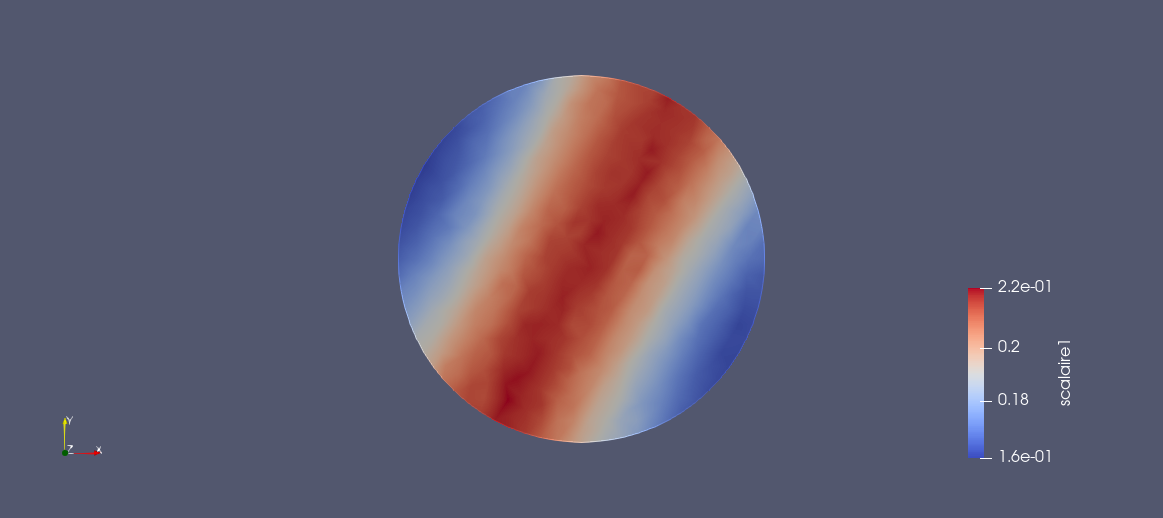}\quad
		\includegraphics[width=0.47\textwidth]{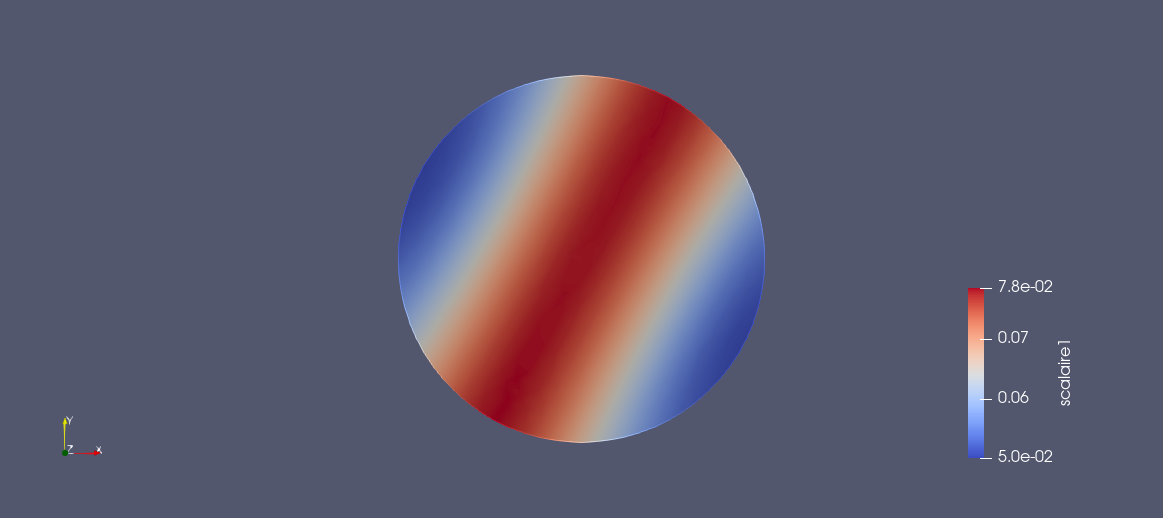}
		\caption{Snapshots of $c_1$ and $c_2$ at timestep $87$. The preferred direction of cell spread is due to the fibre orientation distribution and the
			corresponding haptotaxis of hMSCs towards gradients of $h$ and $\tau$. }
	\end{subfigure}
	\caption{Cell dynamics in the scaffold, 2D simulation}\label{fig:simulation}
\end{figure}

\noindent
Figure \ref{fig:simulation} displays the results of a 2D simulation run for the cell dynamics with a time step $ \Delta t= 0.1$ and the parameters presented in Table \ref{tableparam2} in dimensionless form. 
The stress in the scaffold has been determined from the flow simulation as described in \cite{Grosjean23} and is kept constant here.
As can be seen from the temporal behavior of the different densities evaluated in the midpoint of the scaffold, the hMSCs grow faster than 
the chondrocytes and reach a peak, after which the differentiaton medium $\chi$ is completely consumed. While the concentration of hyaluron 
decays somewhat over time, the production of ECM continues over time. Additionally, the two snapshots of $c_1$ and $c_2$ illustrate the
influence of the orientation distribution that is part of the computation of the diffusion tensor in (\ref{diff-tensor-1}). The cell dynamics is accelerated (due to the taxis term which models migration bias towards gradients of hyaluron - thus of scaffold fibre density and of ECM) along a diagonal line that represents the dominating orientation in the fibres of the scaffold. Since ECM is produced only by chondrocytes and these, in turn, are only obtained by differentiation of hMSCs, the taxis towards $\nabla \tau$ only accentuates the directional bias induced by the structure of the scaffold.\\

%


\begin{minipage}[b]{\textwidth}
	\centering
	\begin{tabular}[h]{|c|c|c|c|}\hline
		{$a_1$} &   $0.015$  &
		$ \beta$& $ 0.5$ \\
		$b_1$& $0.005$ &
		$b_2$ & $0.001$\\
		$\alpha_{min}$ & $0.05$  &
		$\alpha_{max}$ & $0.1$ \\
		$\delta_1$& $0.1 $ &
		$\gamma_1$ &$0.001$  \\
		$ \gamma_2 $ & 0.005 &  $S_{min} $-$ S_{max}$ &$0.1$-$0.3$   \\
		$s_1$ & $30$ &  $s_2$ & $15$  \\
		\hline
	\end{tabular}
	\captionof{table}{Values of model parameters}
	\label{tableparam2}
\end{minipage}

%

\section{Conclusions and perspectives}\label{sec:discussion}



In this note we proposed a multiscale approach to deriving a mathematical model for spread and (de)differentiation of cells involved in tissue regeneration. Their dynamics are (one-way) coupled to fluid flow and scaffold deformation within a bioreactor. Moreover, we also accounted for the evolution of a differentiation medium and of hyaluron, which impregnates the scaffold and is assumed not to diffuse. The effective RDTEs obtained by parabolic upscaling from lower scales are macroscopic, but they carry in their motility terms information about microscopic and mesoscopic dynamics. They appear as a natural consequence of the upscaling process and are not imposed. Moreover, the hMSC diffusion tensor directly encodes the tissue topology. Our macroscopic model obtained here is somehow related to the one in \cite{Pohlmeyer2013}, however extending it in the sense that we perform a more careful characterization of cell dynamics: on the one hand, we consider two cell phenotypes and their transitions; on the other hand, we pay enhanced attention to the anisotropic structure of the fibrous tissue in their surroundings. Instead of the nutrient dynamics considered in \cite{Pohlmeyer2013} we involved the evolution of a growth factor controling (de)differentiation of cells and of a non-diffusing chemical cue (hyaluron) impregnating the scaffold. The simulations show that the cell patterns and thus the newly formed tissue are much influenced by the directional distribution of the scaffold's fibers, via taxis. The latter, in turn, was obtained during the upscaling process. In fact, the analysis performed in \cite{C-MMS} for a simplified version of our macroscopic model shows that cell and tissue patterns are triggered by taxis and not by diffusion.  \\[-2ex]

\noindent
To our knowledge this is the first model for tissue regeneration which includes in an explicit manner the anisotropic topology of the scaffold, statistically assessed from CT data. Furthermore, this is the first approach to simultaneously account for the dynamics of MSCs and their differentiated, matrix-producing counterparts, along with biochemical factors and mechanical effects triggered by fluid flow in a bioreactor and therewith induced deformations of scaffold and ECM. The equations are informed by experimental data. This development was only possible in an interdisciplinary team contributing knowledge from several areas: scaffold production ab´nd cell seeding experiments (biomedical engineering), statistics (data processing), and in silico modeling (mathematics).  \\[-2ex]

\noindent
Our approach combining in vitro experiments with in silico modeling opens the way for investigating a broad palette of questions related to (meniscus) tissue regeneration. Among these, of particular interest are the roles played by geometry, anisotropy, and mechanical properties of the scaffold, along with  phenotype preservation/switch of MSCs seeded in the scaffold under biochemical and biophysical influences. \\[-2ex]

\noindent
The developed models do not only inform the biomedical experiments by suggesting new conjectures and hypotheses, but they also arise interesting mathematical challenges in connection to the analysis (in terms of rigorous convergence of upscaling, well-posedness, patterning, and long term behavior) and numerics of such complex, nonlinear systems coupling equations of several different types and describing processes taking place on different time and space scales.

\appendix

\subsection*{Acknowledgements} This work was funded by the German Research Fundation DFG within SPP 2311. We also acknowledge the support of MathApp at the RPTU Kaiserslautern-Landau. The authors would like to thank Pierre Jolivet (University of Sorbonne, CNRS) for helping with the parallellization of the code, as well as Nishith Mohan and Konstantin Hauch (RPTU Kaiserslautern-Landau) for pointing to us the moment assessment from \cite{ospald} and for image preprocessing, respectively.

\section*{Appendix}

\section{QQ plots for ACG parameter estimation}\label{QQ-plots}

\begin{figure*}
	\begin{subfigure}{\textwidth}
		\centering
		\includegraphics[width=\textwidth]{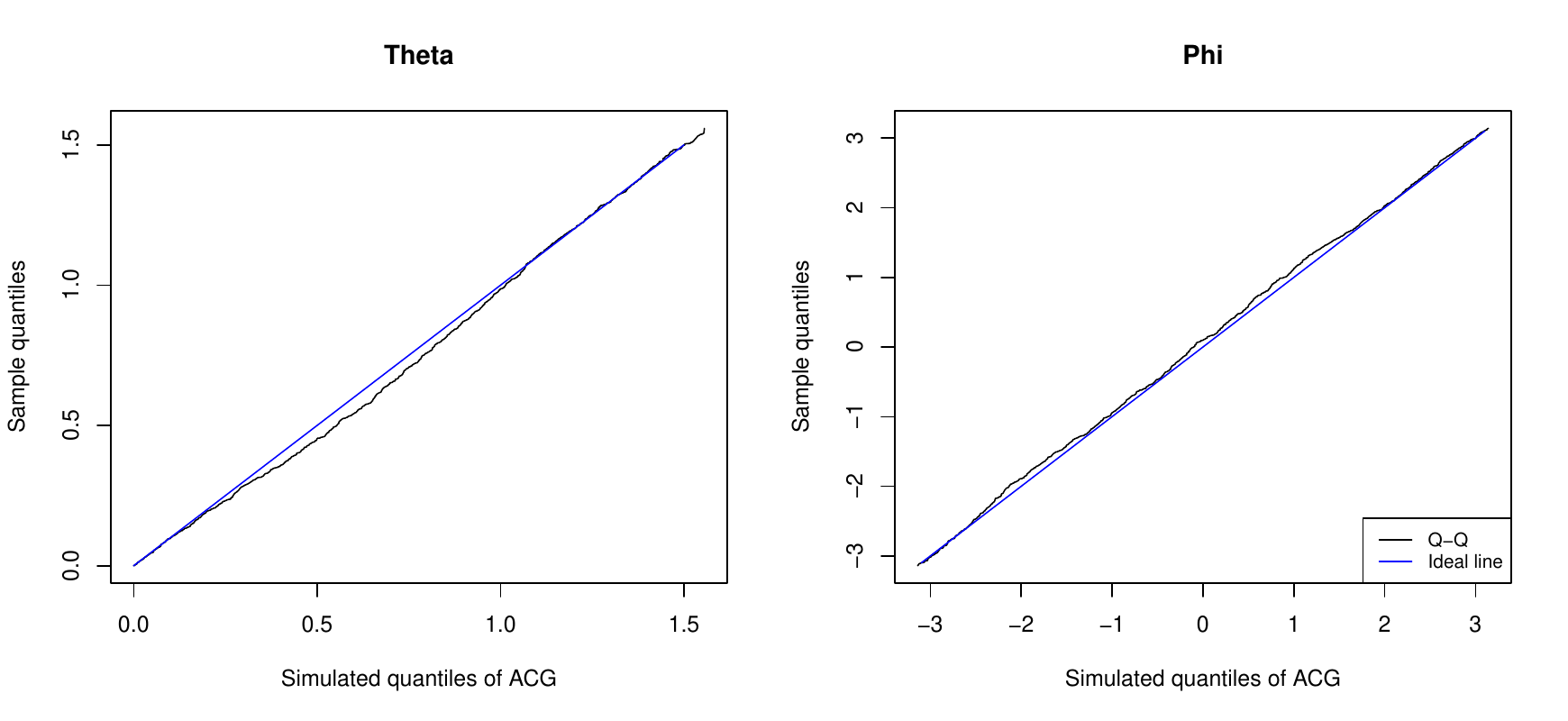}
		\caption{Sample 0.}
		\label{qq-first}
	\end{subfigure}
	\begin{subfigure}{\textwidth}
		\centering
		\includegraphics[width=\textwidth]{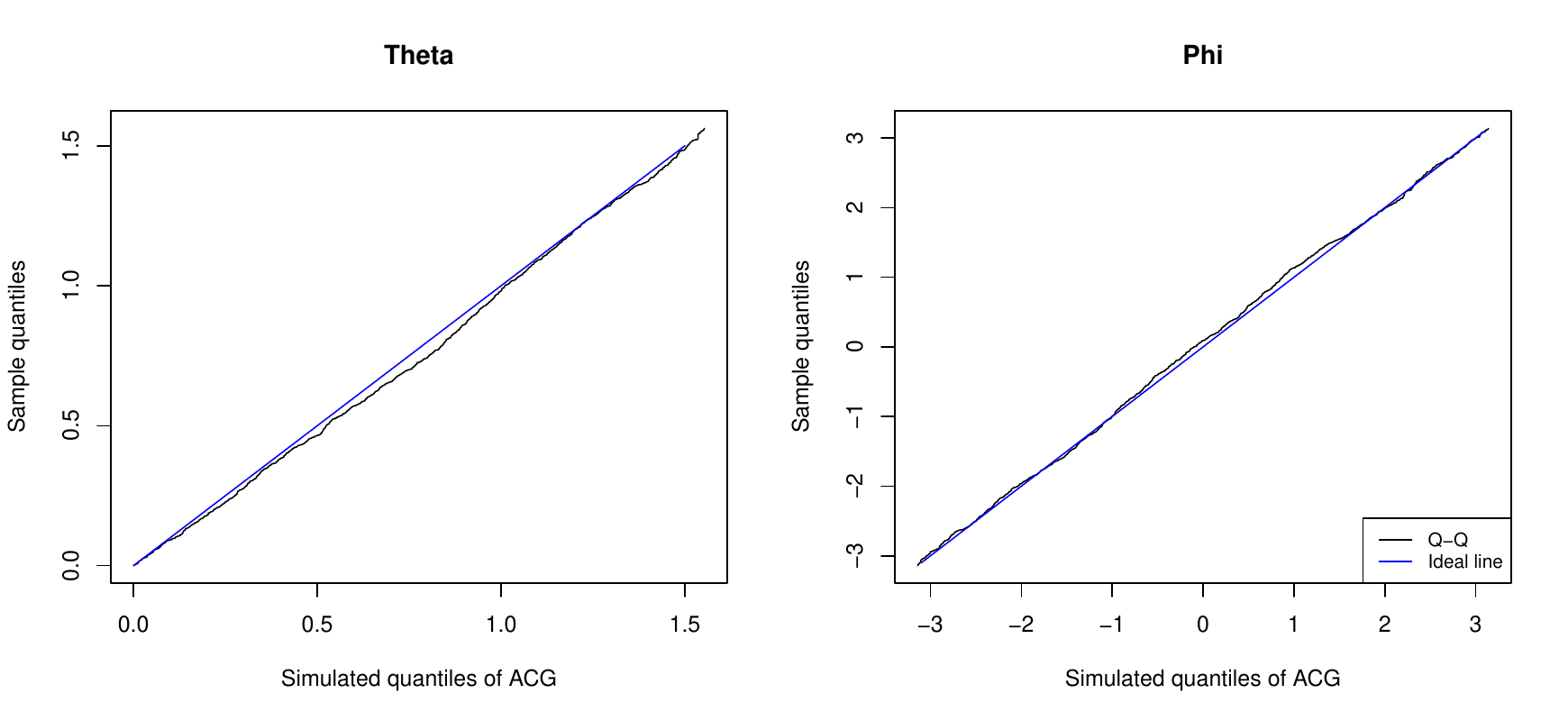}
		\caption{Sample 1.}
	\end{subfigure}
	\begin{subfigure}{\textwidth}
		\centering
		\includegraphics[width=\textwidth]{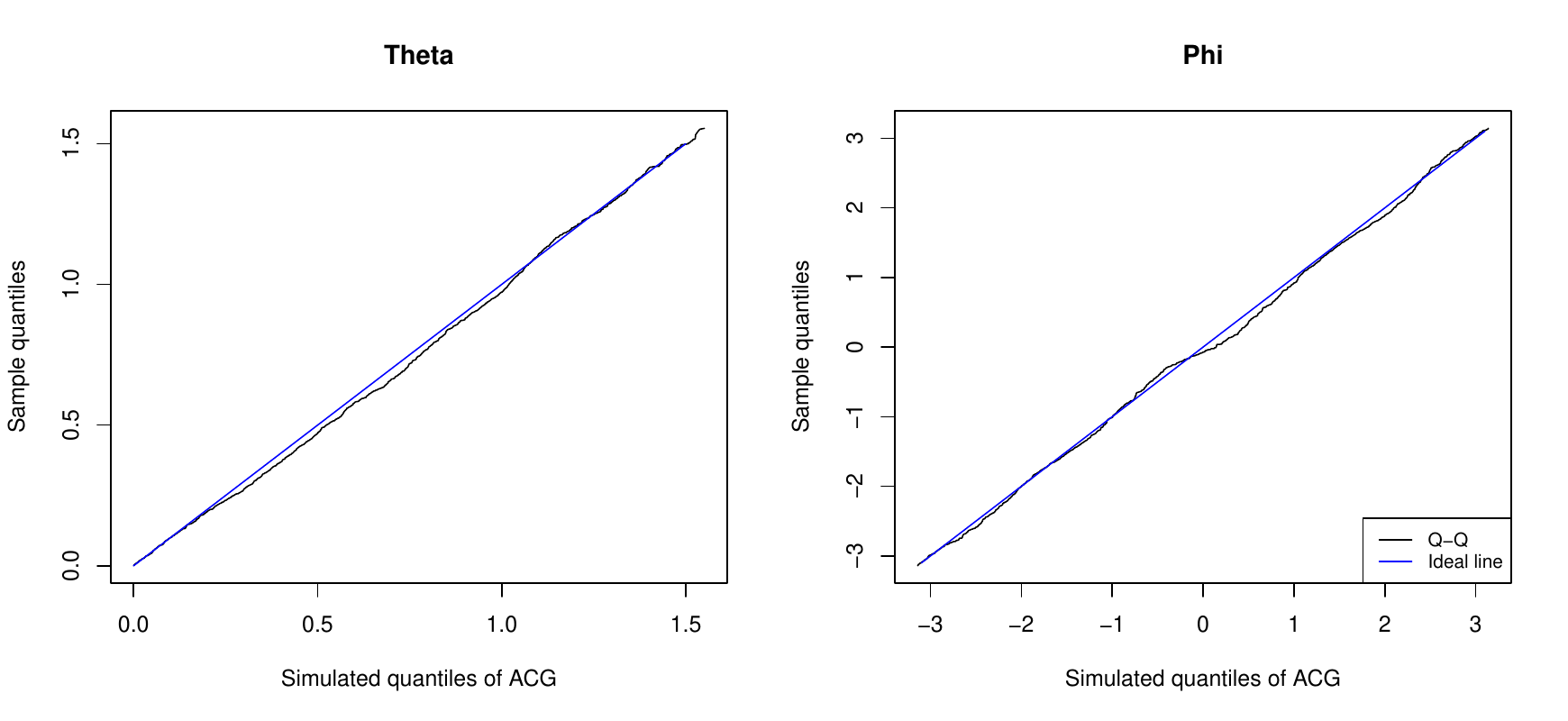}
		\caption{Sample 2.}
	\end{subfigure}
	\caption{QQ plot for the ACG parameter estimation.}
\end{figure*}

\begin{figure*}
	\begin{subfigure}{\textwidth}
		\centering
		\includegraphics[width=\textwidth]{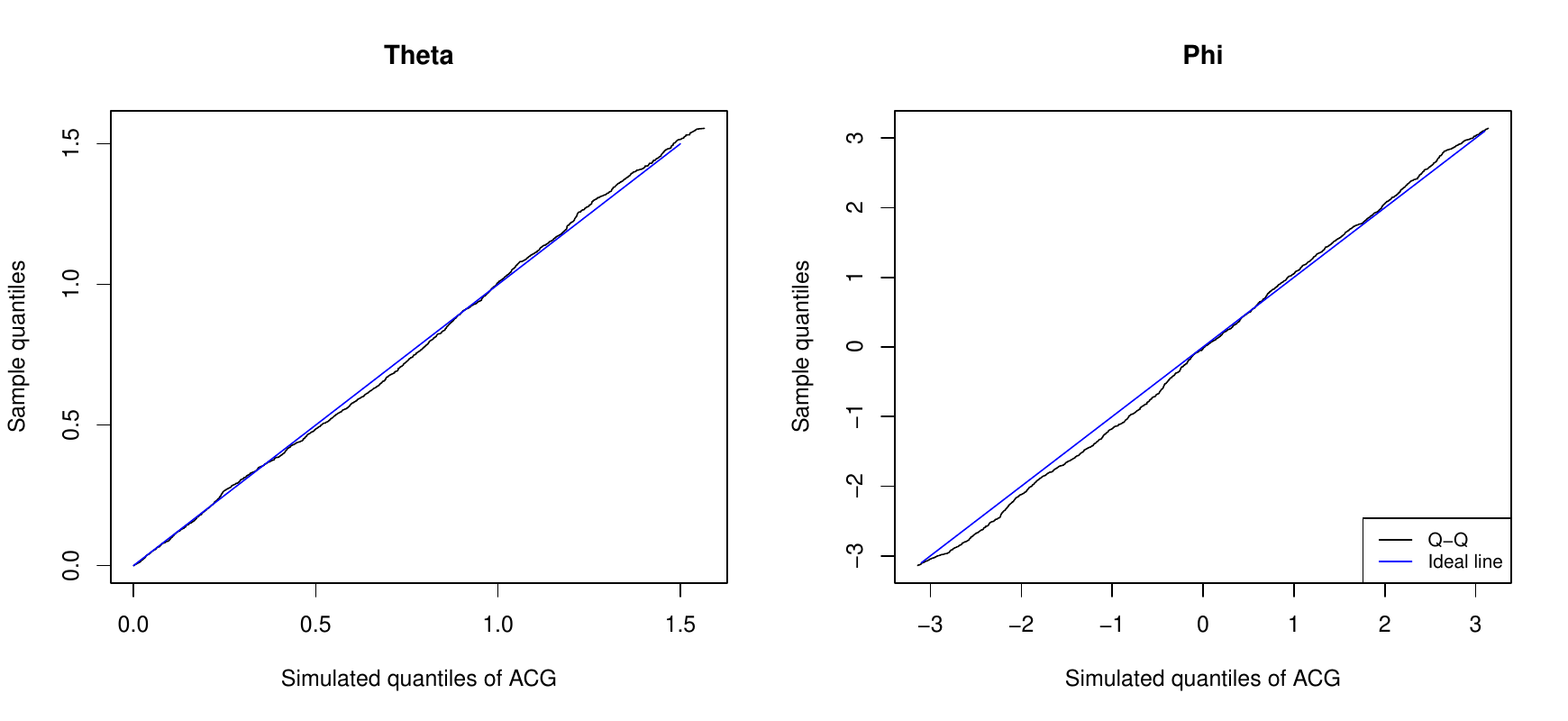}
		\caption{Sample 3.}
	\end{subfigure}
	\begin{subfigure}{\textwidth}
		\centering
		\includegraphics[width=\textwidth]{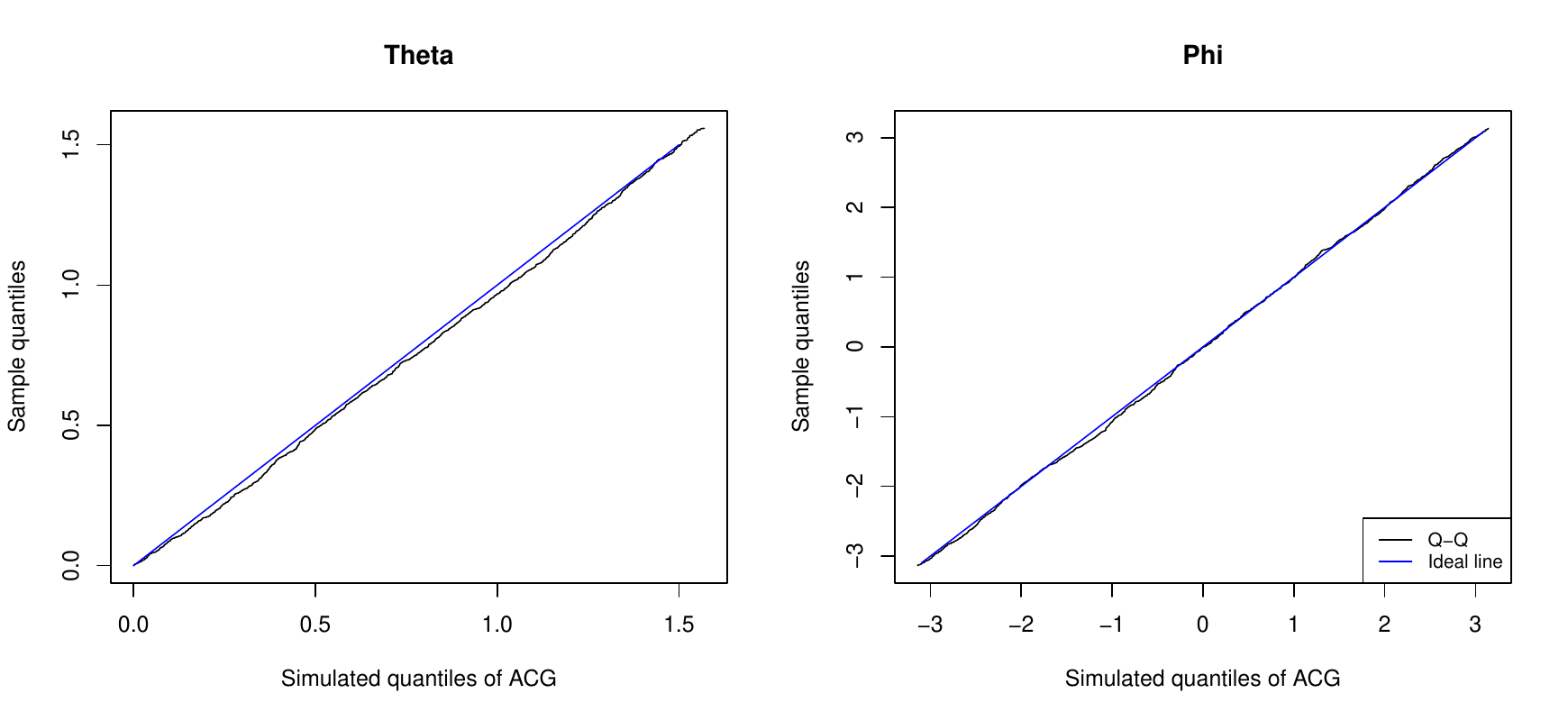}
		\caption{Sample 4.}
	\end{subfigure}
	\begin{subfigure}{\textwidth}
		\centering
		\includegraphics[width=\textwidth]{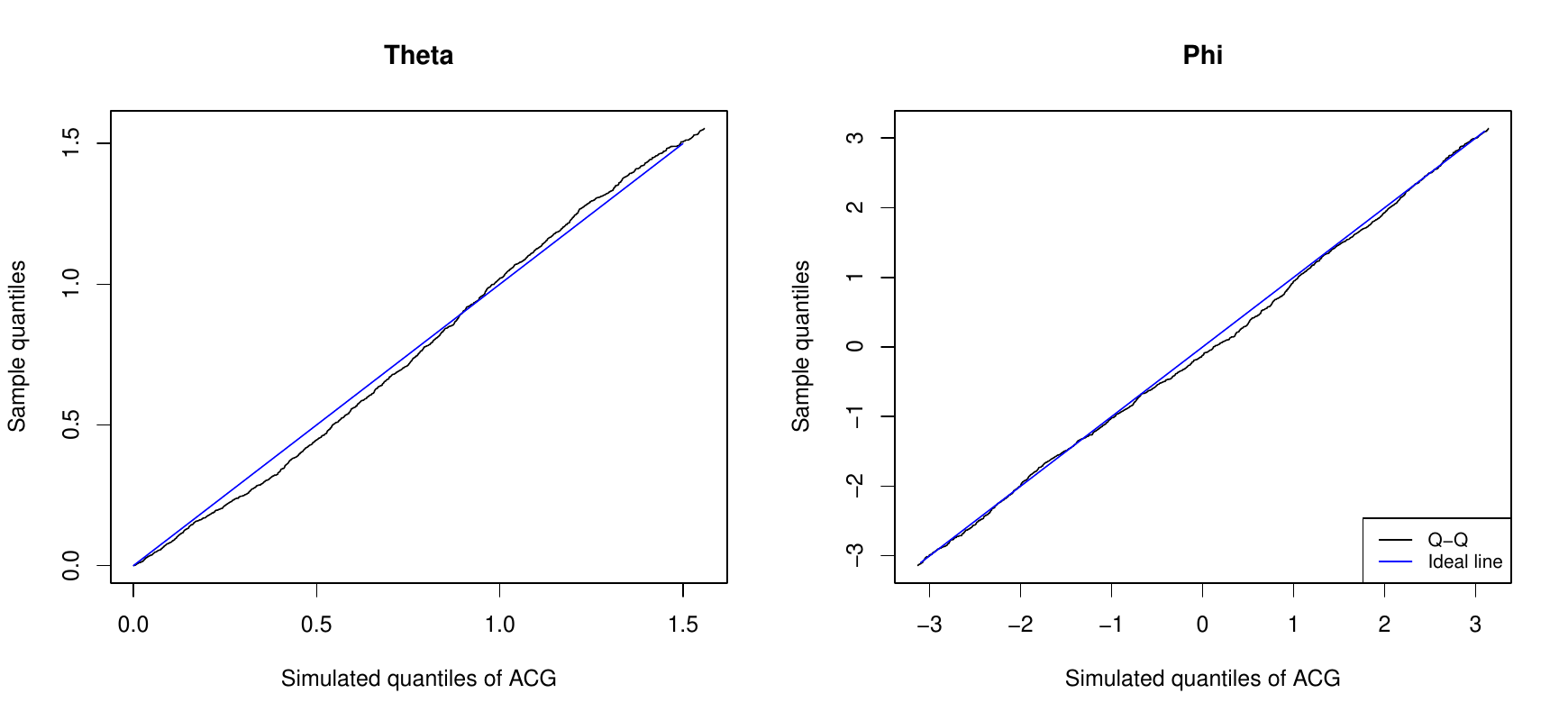}
		\caption{Sample 5.}
	\end{subfigure}
	\caption{QQ plot for the ACG parameter estimation.}
\end{figure*}
\begin{figure*}
	\begin{subfigure}{\textwidth}
		\centering
		\includegraphics[width=\textwidth]{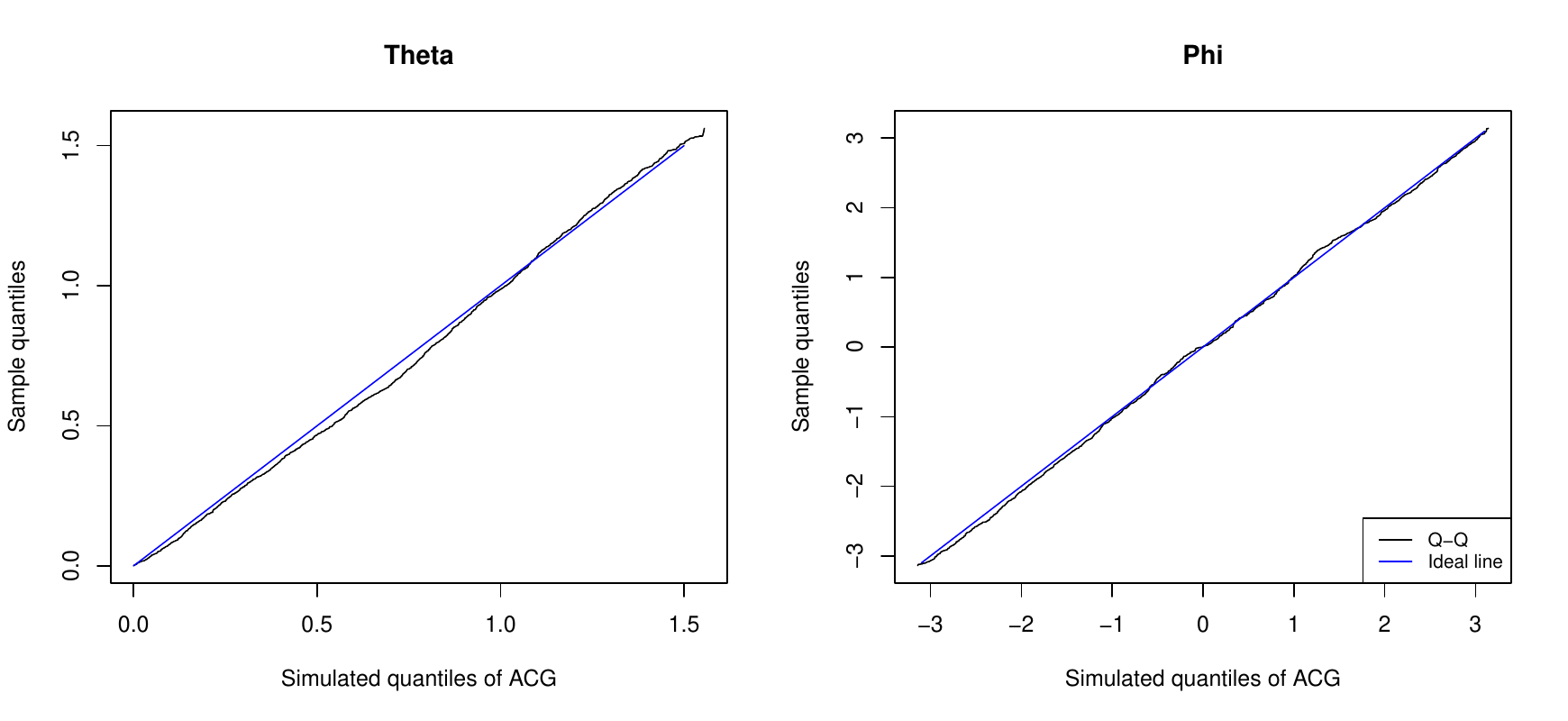}
	\end{subfigure}
	\caption{QQ plot for the ACG parameter estimation of sample 6.}
	\label{qq-last}
\end{figure*}

\clearpage

\phantomsection
\printbibliography

@article{stinner2014global,
  title={Global weak solutions in a PDE-ODE system modeling multiscale cancer cell invasion},
  author={Stinner, Christian and Surulescu, Christina and Winkler, Michael},
  journal={SIAM Journal on Mathematical Analysis},
  volume={46},
  number={3},
  pages={1969--2007},
  year={2014},
  publisher={SIAM}
}

@article{stinner2016global,
  title={Global existence for a go-or-grow multiscale model for tumor invasion with therapy},
  author={Stinner, Christian and Surulescu, Christina and Uatay, Aydar},
  journal={Mathematical Models and Methods in Applied Sciences},
  volume={26},
  number={11},
  pages={2163--2201},
  year={2016},
  publisher={World Scientific}
}

@phdthesis{ospald,
	title={Contributions to the Simulation and Optimization of the Manufacturing Process and the Mechanical Properties of Short Fiber-Reinforced Plastic Parts},
	author={Ospald, Felix},
	year={2019},
	school={TU Chemnitz},
}

@article{altman2002cell,
  title={Cell differentiation by mechanical stress},
  author={Altman, Gregory H and Horan, Rebecca L and Martin, Ivan and Farhadi, Jian and Stark, Peter RH and Volloch, Vladimir and Richmond, John C and Vunjak-Novakovic, Gordana and Kaplan, David L},
  journal={The FASEB Journal},
  volume={16},
  number={2},
  pages={1--13},
  year={2002},
  publisher={Wiley Online Library}
}

@article{park2013control,
  title={Control of cell differentiation by mechanical stress},
  author={Park, Jong-Hoon and Ushida, Takashi and Akimoto, Takayuki},
  journal={The journal of physical fitness and sports medicine},
  volume={2},
  number={1},
  pages={49--62},
  year={2013},
  publisher={The Japanese Society of Physical Fitness and Sports Medicine}
}

@article{mousavi2015role,
  title={Role of mechanical cues in cell differentiation and proliferation: a 3D numerical model},
  author={Mousavi, Seyed Jamaleddin and Hamdy Doweidar, Mohamed},
  journal={PloS one},
  volume={10},
  number={5},
  pages={e0124529},
  year={2015},
  publisher={Public Library of Science San Francisco, CA USA}
}

@article{ghasemi2015structural,
  title={Structural properties of scaffolds: crucial parameters towards stem cells differentiation},
  author={Ghasemi-Mobarakeh, Laleh and Prabhakaran, Molamma P and Tian, Lingling and Shamirzaei-Jeshvaghani, Elham and Dehghani, Leila and Ramakrishna, Seeram},
  journal={World journal of stem cells},
  volume={7},
  number={4},
  pages={728},
  year={2015},
  publisher={Baishideng Publishing Group Inc}
}

@article{freymann2013toward,
  title={Toward scaffold-based meniscus repair: effect of human serum, hyaluronic acid and TGF-{\ss}3 on cell recruitment and re-differentiation},
  author={Freymann, U and Endres, M and Goldmann, U and Sittinger, M and Kaps, C},
  journal={Osteoarthritis and Cartilage},
  volume={21},
  number={5},
  pages={773--781},
  year={2013},
  publisher={Elsevier}
}

@article{mauck2007regional,
  title={Regional multilineage differentiation potential of meniscal fibrochondrocytes: implications for meniscus repair},
  author={Mauck, Robert L and Martinez-Diaz, Gabriel J and Yuan, Xiaoning and Tuan, Rocky S},
  journal={The Anatomical Record: Advances in Integrative Anatomy and Evolutionary Biology: Advances in Integrative Anatomy and Evolutionary Biology},
  volume={290},
  number={1},
  pages={48--58},
  year={2007},
  publisher={Wiley Online Library}
}

@article{mishima2008chemotaxis,
  title={Chemotaxis of human articular chondrocytes and mesenchymal stem cells},
  author={Mishima, Yasunori and Lotz, Martin},
  journal={Journal of Orthopaedic Research},
  volume={26},
  number={10},
  pages={1407--1412},
  year={2008},
  publisher={Wiley Online Library}
}

@article{geris2008angiogenesis,
  title={Angiogenesis in bone fracture healing: a bioregulatory model},
  author={Geris, Liesbet and Gerisch, Alf and Vander Sloten, Jos and Weiner, R{\"u}diger and Van Oosterwyck, Hans},
  journal={Journal of theoretical biology},
  volume={251},
  number={1},
  pages={137--158},
  year={2008},
  publisher={Elsevier}
}

@article{bellomo2015toward,
  title={Toward a mathematical theory of Keller--Segel models of pattern formation in biological tissues},
  author={Bellomo, Nicola and Bellouquid, Abdelghani and Tao, Youshan and Winkler, Michael},
  journal={Mathematical Models and Methods in Applied Sciences},
  volume={25},
  number={09},
  pages={1663--1763},
  year={2015},
  publisher={World Scientific}
}

@article{kolbe2021modeling,
  title={Modeling multiple taxis: tumor invasion with phenotypic heterogeneity, haptotaxis, and unilateral interspecies repellence},
  author={Kolbe, Niklas and Sfakianakis, Nikolaos and Stinner, Christian and Surulescu, Christina and Lenz, Jonas},
  journal={Discr. Cont. Dyn. Syst. B},
  volume={26},
  year={2021},
  pages={443-481},
}

@article{zhigun2018strongly,
  title={A strongly degenerate diffusion-haptotaxis model of tumour invasion under the go-or-grow dichotomy hypothesis},
  author={Zhigun, Anna and Surulescu, Christina and Hunt, Alexander},
  journal={Mathematical Methods in the Applied Sciences},
  volume={41},
  number={6},
  pages={2403--2428},
  year={2018},
  publisher={Wiley Online Library}
}

@article{zhigun2016global,
  title={Global existence for a degenerate haptotaxis model of cancer invasion},
  author={Zhigun, Anna and Surulescu, Christina and Uatay, Aydar},
  journal={Zeitschrift f{\"u}r angewandte Mathematik und Physik},
  volume={67},
  pages={1--29},
  year={2016},
  publisher={Springer}
}

@article{eckardt2020nonlocal,
  title={Nonlocal and local models for taxis in cell migration: a rigorous limit procedure},
  author={Eckardt, Maria and Painter, Kevin J and Surulescu, Christina and Zhigun, Anna},
  journal={Journal of Mathematical Biology},
  volume={81},
  pages={1251--1298},
  year={2020},
  publisher={Springer}
}

@article{barocas1997anisotropic,
  title={An anisotropic biphasic theory of tissue-equivalent mechanics: the interplay among cell traction, fibrillar network deformation, fibril alignment, and cell contact guidance},
  author={Barocas, Victor H and Tranquillo, Robert T},
  year={1997}
}

@article{lemon2007travelling,
  title={Travelling-wave behaviour in a multiphase model of a population of cells in an artificial scaffold},
  author={Lemon, G and King, JR},
  journal={Journal of Mathematical Biology},
  volume={55},
  pages={449--480},
  year={2007},
  publisher={Springer}
}

@article{klika2016overview,
  title={An overview of multiphase cartilage mechanical modelling and its role in understanding function and pathology},
  author={Klika, Vaclav and Gaffney, Eamonn A and Chen, Ying-Chun and Brown, Cameron P},
  journal={journal of the mechanical behavior of biomedical materials},
  volume={62},
  pages={139--157},
  year={2016},
  publisher={Elsevier}
}

@article{jackson2002mechanical,
  title={A mechanical model of tumor encapsulation and transcapsular spread},
  author={Jackson, Trachette L and Byrne, Helen M},
  journal={Mathematical biosciences},
  volume={180},
  number={1-2},
  pages={307--328},
  year={2002},
  publisher={Elsevier}
}

@article{kumar2021multiphase,
  title={Multiphase modelling of glioma pseudopalisading under acidosis},
  author={Kumar, Pawan and Surulescu, Christina and Zhigun, Anna},
  journal={arXiv preprint arXiv:2106.15241},
  year={2021}
}

@book{bellomo2017quest,
  title={A quest towards a mathematical theory of living systems},
  author={Bellomo, Nicola and Bellouquid, Abdelghani and Gibelli, Livio and Outada, Nisrine},
  year={2017},
  publisher={Springer}
}

@article{conte2023mathematical,
  title={Mathematical modeling of glioma invasion and therapy approaches via kinetic theory of active particles},
  author={Conte, Martina and Dzierma, Yvonne and Knobe, Sven and Surulescu, Christina},
  journal={Mathematical Models and Methods in Applied Sciences},
  pages={1--43},
  year={2023},
  publisher={World Scientific}
}

@article{corbin2021modeling,
  title={Modeling glioma invasion with anisotropy-and hypoxia-triggered motility enhancement: From subcellular dynamics to macroscopic PDEs with multiple taxis},
  author={Corbin, Gregor and Klar, Axel and Surulescu, Christina and Engwer, Christian and Wenske, Michael and Nieto, Juanjo and Soler, Juan},
  journal={Mathematical Models and Methods in Applied Sciences},
  volume={31},
  number={01},
  pages={177--222},
  year={2021},
  publisher={World Scientific}
}

@article{dietrich2022multiscale,
  title={Multiscale modeling of glioma invasion: from receptor binding to flux-limited macroscopic PDEs},
  author={Dietrich, Anne and Kolbe, Niklas and Sfakianakis, Nikolaos and Surulescu, Christina},
  journal={Multiscale Modeling \& Simulation},
  volume={20},
  number={2},
  pages={685--713},
  year={2022},
  publisher={SIAM}
}

@article{engwer2015glioma,
  title={Glioma follow white matter tracts: a multiscale DTI-based model},
  author={Engwer, Christian and Hillen, Thomas and Knappitsch, Markus and Surulescu, Christina},
  journal={Journal of mathematical biology},
  volume={71},
  pages={551--582},
  year={2015},
  publisher={Springer}
}

@article{engwer2016effective,
  title={Effective equations for anisotropic glioma spread with proliferation: a multiscale approach and comparisons with previous settings},
  author={Engwer, Christian and Hunt, Alexander and Surulescu, Christina},
  journal={Mathematical medicine and biology: a journal of the IMA},
  volume={33},
  number={4},
  pages={435--459},
  year={2016},
  publisher={Oxford University Press}
}

@article{kumar2021multiscale,
  title={Multiscale modeling of glioma pseudopalisades: contributions from the tumor microenvironment},
  author={Kumar, Pawan and Li, Jing and Surulescu, Christina},
  journal={Journal of Mathematical Biology},
  volume={82},
  pages={1--45},
  year={2021},
  publisher={Springer}
}

@article{painter2013mathematical,
  title={Mathematical modelling of glioma growth: the use of diffusion tensor imaging (DTI) data to predict the anisotropic pathways of cancer invasion},
  author={Painter, KJ and Hillen, Thomas},
  journal={Journal of theoretical biology},
  volume={323},
  pages={25--39},
  year={2013},
  publisher={Elsevier}
}

@article{zhigun2022novel,
  title={A novel derivation of rigorous macroscopic limits from a micro-meso description of signal-triggered cell migration in fibrous environments},
  author={Zhigun, Anna and Surulescu, Christina},
  journal={SIAM Journal on Applied Mathematics},
  volume={82},
  number={1},
  pages={142--167},
  year={2022},
  publisher={SIAM}
}

@article{conte2021mathematical,
  title={Mathematical modeling of glioma invasion: acid-and vasculature mediated go-or-grow dichotomy and the influence of tissue anisotropy},
  author={Conte, Martina and Surulescu, Christina},
  journal={Applied Mathematics and Computation},
  volume={407},
  pages={126305},
  year={2021},
  publisher={Elsevier}
}

@article{aufderheide2004mechanical,
  title={Mechanical stimulation toward tissue engineering of the knee meniscus},
  author={AufderHeide, Adam C and Athanasiou, Kyriacos A},
  journal={Annals of biomedical engineering},
  volume={32},
  pages={1163--1176},
  year={2004},
  publisher={Springer}
}

@article{fahy2018mechanical,
  title={Mechanical stimulation of mesenchymal stem cells: Implications for cartilage tissue engineering},
  author={Fahy, Niamh and Alini, Mauro and Stoddart, Martin J},
  journal={Journal of Orthopaedic Research},
  volume={36},
  number={1},
  pages={52--63},
  year={2018},
  publisher={Wiley Online Library}
}

@article{janmey2007cell,
  title={Cell mechanics: integrating cell responses to mechanical stimuli},
  author={Janmey, Paul A and McCulloch, Christopher A},
  journal={Annu. Rev. Biomed. Eng.},
  volume={9},
  pages={1--34},
  year={2007},
  publisher={Annual Reviews}
}

@article{lee2006influence,
  title={The influence of mechanical stimuli on articular cartilage tissue engineering},
  author={Lee, C and Grad, S and Wimmer, M and Alini, M and others},
  journal={Topics in tissue engineering},
  volume={2},
  number={2},
  year={2006},
  publisher={Expertissues Oulu, Finland}
}

@article{huiskes1997biomechanical,
  title={A biomechanical regulatory model for periprosthetic fibrous-tissue differentiation},
  author={Huiskes, Rik and Driel, WD Van and Prendergast, Patrick J and S{\o}balle, Kjeld and others},
  journal={Journal of materials science: Materials in medicine},
  volume={8},
  number={12},
  pages={785--788},
  year={1997},
  publisher={Springer-Verlag, Tiergartenstrasse 17 Heidelberg 69121 Germany}
}

@article{andreykiv2008simulation,
  title={Simulation of fracture healing incorporating mechanoregulation of tissue differentiation and dispersal/proliferation of cells},
  author={Andreykiv, A and Van Keulen, F and Prendergast, PJ},
  journal={Biomechanics and modeling in mechanobiology},
  volume={7},
  pages={443--461},
  year={2008},
  publisher={Springer}
}

@article{Lorenz2014,
	doi = {10.1142/s0218202514500249},
	url = {https://doi.org/10.1142/s0218202514500249},
	year = {2014},
	month = aug,
	publisher = {World Scientific Pub Co Pte Lt},
	volume = {24},
	number = {12},
	pages = {2383--2436},
	author = {T. Lorenz and C. Surulescu},
	title = {On a class of multiscale cancer cell migration models: Well-posedness in less regular function spaces},
	journal = {Mathematical Models and Methods in Applied Sciences}
}

@article{OthmerHillen2000,
	title={The Diffusion Limit of Transport Equations Derived from Velocity-Jump Processes},
	author={Hans G. Othmer and Thomas Hillen},
	journal={SIAM J. Appl. Math.},
	year={2000},
	volume={61},
	pages={751-775}
}

@article{plaza,
	doi = {10.1007/s00285-018-1323-x},
	url = {https://doi.org/10.1007/s00285-018-1323-x},
	year = {2019},
	publisher = {Springer Science and Business Media {LLC}},
	volume = {78},
	number = {6},
	pages = {1681--1711},
	author = {Ram{\'{o}}n G. Plaza},
	title = {Derivation of a bacterial nutrient-taxis system with doubly degenerate cross-diffusion as the parabolic limit of a velocity-jump process},
	journal = {Journal of Mathematical Biology}
}

@incollection{AP08,
	author = {Astanin, Sergey and Preziosi, Luigi},
	booktitle = {Selected Topics in Cancer Modeling: Genesis - Evolution - Immune Competition - Therapy},
	editor = {N. Bellomo and M.A.J. Chaplain and E. De Angelis},
	pages = {1-31},
	publisher = {Springer},
	title = {Multiphase Models of Tumour Growth},
	year = {2008},
	doi = {10.1007/978-0-8176-4713-1_9},
}

@article{Evje-Winkler20,
	doi = {10.1007/s00332-020-09625-w},
	url = {https://doi.org/10.1007/s00332-020-09625-w},
	year = {2020},
	publisher = {Springer Science and Business Media {LLC}},
	volume = {30},
	number = {4},
	pages = {1809--1847},
	author = {Steinar Evje and Michael Winkler},
	title = {Mathematical Analysis of Two Competing Cancer Cell Migration Mechanisms Driven by Interstitial Fluid Flow},
	journal = {Journal of Nonlinear Science}
}

@article{Prendergast97,
	doi = {10.1016/s0021-9290(96)00140-6},
	url = {https://doi.org/10.1016/s0021-9290(96)00140-6},
	year = {1997},
	publisher = {Elsevier {BV}},
	volume = {30},
	number = {6},
	pages = {539--548},
	author = {P.J. Prendergast and R. Huiskes and K. S{\o}balle},
	title = {Biophysical stimuli on cells during tissue differentiation at implant interfaces},
	journal = {Journal of Biomechanics}
}

@article{Grosjean23,
	title = {A mathematical model for meniscus cartilage regeneration},
	volume = {23},
	ISSN = {1617-7061},
	url = {http://dx.doi.org/10.1002/pamm.202300261},
	DOI = {10.1002/pamm.202300261},
	number = {3},
	journal = {PAMM},
	publisher = {Wiley},
	author = {Grosjean,  E. and Simeon,  B. and Surulescu,  C.},
	year = {2023},
}

@article{Ambarts,
doi = {10.1007/s00211-018-0967-1},
url = {https://doi.org/10.1007/s00211-018-0967-1},
year = {2018},
publisher = {Springer Science and Business Media {LLC}},
volume = {140},
number = {2},
pages = {513-553},
author = {Ilona Ambartsumyan and Eldar Khattatov and Ivan Yotov and Paolo Zunino},
title = {A Lagrange multiplier method for a Stokes{\textendash}Biot fluid{\textendash}poroelastic structure interaction model},
journal = {Numerische Mathematik},
}

@article{biotdarcystokes1,
	doi = {10.1016/j.cma.2014.10.047},
	url = {https://doi.org/10.1016/j.cma.2014.10.047},
	year = {2015},
	publisher = {Elsevier {BV}},
	volume = {292},
	pages = {138-170},
	author = {M. Buka{\v{c}} and I. Yotov and R. Zakerzadeh and P. Zunino},
	title = {Partitioning strategies for the interaction of a fluid with a poroelastic material based on a Nitsche's coupling approach},
	journal = {Computer Methods in Applied Mechanics and Engineering}
}

@Article{otsu79,
	author =       "N. Otsu",
	title =        "A threshold selection method from gray level
	histograms",
	journal =      "{IEEE} Trans. Systems, Man and Cybernetics",
	year =         "1979",
	volume =       "9",
	pages =        "62--66",
	month =        mar
}

@book{fisher1987StatisticalAnalysisSpherical,
	title = {Statistical Analysis of Spherical Data},
	author = {Fisher, N. I. and Lewis, Toby and Embleton, B. J. J.},
	date = {1987},
	publisher = {{Cambridge University Press}},
	location = {{Cambridge [Cambridgeshire] ; New York}},
	isbn = {978-0-521-24273-8},
	langid = {english},
	pagetotal = {329},
}

@article{ref:ulm1,
	doi = {10.3389/fbioe.2021.659989},
	url = {https://doi.org/10.3389/fbioe.2021.659989},
	year = {2021},
	publisher = {Frontiers Media {SA}},
	volume = {9},
	author = {Andreas M. Seitz and Felix Osthaus and Jonas Schwer and Daniela Warnecke and Martin Faschingbauer and Mirco Sgroi and Anita Ignatius and Lutz D\"{u}rselen},
	title = {Osteoarthritis-Related Degeneration Alters the Biomechanical Properties of Human Menisci Before the Articular Cartilage},
	journal = {Frontiers in Bioengineering and Biotechnology}
}

@article{ref:ulm2,
	doi = {10.1002/jor.23330},
	url = {https://doi.org/10.1002/jor.23330},
	year = {2016},
	publisher = {Wiley},
	volume = {35},
	number = {4},
	pages = {858--867},
	author = {Sotcheadt Sim and Anik Chevrier and Martin Garon and Eric Quenneville and Patrick Lavigne and Alex Yaroshinsky and Caroline D. Hoemann and Michael D. Buschmann},
	title = {Electromechanical probe and automated indentation maps are sensitive techniques in assessing early degenerated human articular cartilage},
	journal = {Journal of Orthopaedic Research}
}

@article{ref:ulm3,
	doi = {10.1016/0021-9290(72)90010-3},
	url = {https://doi.org/10.1016/0021-9290(72)90010-3},
	year = {1972},
	month = sep,
	publisher = {Elsevier {BV}},
	volume = {5},
	number = {5},
	pages = {541--551},
	author = {W.C. Hayes and L.M. Keer and G. Herrmann and L.F. Mockros},
	title = {A mathematical analysis for indentation tests of articular cartilage},
	journal = {Journal of Biomechanics}
}

@article{ref:ulm4,
	doi = {10.1016/j.jmbbm.2013.05.027},
	url = {https://doi.org/10.1016/j.jmbbm.2013.05.027},
	year = {2013},
	publisher = {Elsevier {BV}},
	volume = {26},
	pages = {68--80},
	author = {Andreas Martin Seitz and Fabio Galbusera and Carina Krais and Anita Ignatius and Lutz D\"{u}rselen},
	title = {Stress-relaxation response of human menisci under confined compression conditions},
	journal = {Journal of the Mechanical Behavior of Biomedical Materials}
}

@article{ref:ulm5,
	doi = {10.1002/adhm.202301787},
	url = {https://doi.org/10.1002/adhm.202301787},
	year = {2023},
	publisher = {Wiley},
	author = {Muthusamy Saranya and Aldeliane M. da Silva and Hanna Karjalainen and Geir Klinkenberg and Ruth Schmid and Birgitte McDonagh and Peter P. Molesworth and Margr{\'{e}}t S. Sigf{\'{u}}sd{\'{o}}ttir and Ane Marit W{\aa}gb{\o} and Susana. G. Santos and Cristiana Couto and Ville-Pauli Karjalainen and Shuvashis Das Gupta and Topias J\"{a}rvinen and Luisa de Roy and Andreas. M. Seitz and Mikko Finnil\"{a} and Simo Saarakkala and Anne Marie Haaparanta and Lauriane Janssen and Gabriela S. Lorite},
	title = {Magnetic-Responsive Carbon Nanotubes Composite Scaffolds for Chondrogenic Tissue Engineering},
	journal = {Advanced Healthcare Materials}
}

@incollection{ref:ulm6,
author = {Van C. Mow and R. Huiskes},
booktitle = {Basic Orthopaedic Biomechanics \& Mechano-Biology.},
editor = {Van C. Mow and Weiyong Gu and F. H. Chen},
pages = {182-257},
publisher = {Philadelphia: Lippincott Williams \& Wilkins},
title = {Structure and function of articular cartilage and meniscus},
year = {2005},
}

@Book{KMS95,
	author =       {S. Chiu and D. Stoyan and W. Kendall and J. Mecke},
	year =         {2013},
	title =        {Stochastic geometry and its applications},
	publisher =    {John Wiley \& sons},
}

@article{tyler1987StatisticalAnalysisAngular,
	doi = {10.1093/biomet/74.3.579},
	url = {https://doi.org/10.1093/biomet/74.3.579},
	year = {1987},
	publisher = {Oxford University Press ({OUP})},
	volume = {74},
	number = {3},
	pages = {579--589},
	author = {David E. Tyler},
	title = {Statistical analysis for the angular central Gaussian distribution on the sphere},
	journal = {Biometrika}
}

@Manual{Rfast,
	title = {Rfast: A Collection of Efficient and Extremely Fast R Functions},
	author = {Manos Papadakis and Michail Tsagris and Marios Dimitriadis and Stefanos Fafalios and Ioannis Tsamardinos and Matteo Fasiolo and Giorgos Borboudakis and John Burkardt and Changliang Zou and Kleanthi Lakiotaki and Christina Chatzipantsiou.},
	year = {2023},
	note = {R package version 2.0.8},
	url = {https://CRAN.R-project.org/package=Rfast},
}

@Article{wirjadi16,
	author = 	 {Wirjadi, O. and Schladitz, K. and Easwaran, P. and Ohser, J.},
	title = 	 {Estimating Fibre Direction Distributions of Reinforced Composites from Tomographic Images},
	journal = 	 {Image Anal. Stereol.},
	year = 	 2016,
	volume =       35,
	number =       3,
	pages =        {167-179 \url{http://dx.doi.org/10.5566/ias.1489}},
	url = 	 {<https://www.ias-iss.org/ojs/IAS/article/view/1489>}, 
	doi =          {http://dx.doi.org/10.5566/ias.1489}
}

@article{eberly94,
	journal = "J. Mathematical Imaging and Vision",
	title = "Ridges for image analysis",
	volume = 4,
	number = 4,
	pages = "353-373. \url{https://doi.org/10.1007/BF01262402}",
	year = 1994,
	author = "D. Eberly and R. Gardner and B. Morse and S. Pizer and C. Scharlach",
	doi = "10.1007/BF01262402"
}

@Book{ohser-schladitz09book,
	author =	 {Ohser, J. and Schladitz, K.},
	title = 	 {3d Images of Materials Structures -- Processing and Analysis},
	publisher = 	 {Wiley VCH},
	year = 	 2009,
	address =      {Weinheim}
}

@article{pinter18,
	author = {Pinter, Pascal and Dietrich, Stefan and Bertram, B. and Kehrer, Loredana and Elsner, P. and Weidenmann, K.A.},
	year = {2018},
	pages = {},
	title = {Comparison and error estimation of 3D fibre orientation analysis of computed tomography image data for fibre reinforced composites},
	volume = {95. \url{https://doi.org/10.1016/j.ndteint.2018.01.001}},
	journal = {NDT~\&~E Int.},
	doi = {10.1016/j.ndteint.2018.01.001}
}

@book{di2011mathematical,
	doi = {10.1007/978-3-642-22980-0},
	url = {https://doi.org/10.1007/978-3-642-22980-0},
	year = {2012},
	publisher = {Springer Berlin Heidelberg},
	author = {Daniele Antonio Di Pietro and Alexandre Ern},
	title = {Mathematical Aspects of Discontinuous Galerkin Methods}
}

@article{BAILONPLAZA2001,
	title = {A Mathematical Framework to Study the Effects of Growth Factor Influences on Fracture Healing},
	volume = {212},
	ISSN = {0022-5193},
	url = {http://dx.doi.org/10.1006/jtbi.2001.2372},
	DOI = {10.1006/jtbi.2001.2372},
	number = {2},
	journal = {Journal of Theoretical Biology},
	publisher = {Elsevier BV},
	author = {Bail\'on-Plaza,  Alicia and Van der Meulen,  Marjolein C.H.},
	year = {2001},
	pages = {191-209}
}

@article{GmezBenito2005,
	title = {Influence of fracture gap size on the pattern of long bone healing: a computational study},
	volume = {235},
	ISSN = {0022-5193},
	url = {http://dx.doi.org/10.1016/j.jtbi.2004.12.023},
	DOI = {10.1016/j.jtbi.2004.12.023},
	number = {1},
	journal = {Journal of Theoretical Biology},
	publisher = {Elsevier BV},
	author = {Gómez-Benito,  M.J. and García-Aznar,  J.M. and Kuiper,  J.H. and Doblaré,  M.},
	year = {2005},
	pages = {105–119}
}

@article{Ribeiro2015,
	title = {In silico Mechano-Chemical Model of Bone Healing for the Regeneration of Critical Defects: The Effect of BMP-2},
	volume = {10},
	ISSN = {1932-6203},
	url = {http://dx.doi.org/10.1371/journal.pone.0127722},
	DOI = {10.1371/journal.pone.0127722},
	number = {6},
	journal = {PLOS ONE},
	publisher = {Public Library of Science (PLoS)},
	author = {Ribeiro,  Frederico O. and Gómez-Benito,  María José and Folgado,  João and Fernandes,  Paulo R. and García-Aznar,  José Manuel},
	editor = {Yamamoto,  Masaya},
	year = {2015},
	pages = {e0127722}
}

@article{Campbell2019,
	title = {A mathematical model of cartilage regeneration after chondrocyte and stem cell implantation – I: the effects of growth factors},
	volume = {10},
	ISSN = {2041-7314},
	url = {http://dx.doi.org/10.1177/2041731419827791},
	DOI = {10.1177/2041731419827791},
	journal = {Journal of Tissue Engineering},
	publisher = {SAGE Publications},
	author = {Campbell,  Kelly and Naire,  Shailesh and Kuiper,  Jan Herman},
	year = {2019},
	pages = {204173141982779}
}

@article{Campbell2019-b,
	title = {A mathematical model of cartilage regeneration after chondrocyte and stem cell implantation – II: the effects of co-implantation},
	volume = {10},
	ISSN = {2041-7314},
	url = {http://dx.doi.org/10.1177/2041731419827792},
	DOI = {10.1177/2041731419827792},
	journal = {Journal of Tissue Engineering},
	publisher = {SAGE Publications},
	author = {Campbell,  Kelly and Naire,  Shailesh and Kuiper,  Jan Herman},
	year = {2019},
	pages = {204173141982779}
}

@article{WSE-H,
	title = {Regenerative medicine meets mathematical modelling: developing symbiotic relationships},
	volume = {6},
	ISSN = {2057-3995},
	url = {http://dx.doi.org/10.1038/s41536-021-00134-2},
	DOI = {10.1038/s41536-021-00134-2},
	number = {1},
	journal = {npj Regenerative Medicine},
	publisher = {Springer Science and Business Media LLC},
	author = {Waters,  S. L. and Schumacher,  L. J. and El Haj,  A. J.},
	year = {2021},
}

@article{CPSZ,
title = {Mathematical models for cell migration: a non-local perspective},
volume = {375},
ISSN = {1471-2970},
url = {http://dx.doi.org/10.1098/rstb.2019.0379},
DOI = {10.1098/rstb.2019.0379},
number = {1807},
journal = {Philosophical Transactions of the Royal Society B: Biological Sciences},
publisher = {The Royal Society},
author = {Chen,  Li and Painter,  Kevin and Surulescu,  Christina and Zhigun,  Anna},
year = {2020},
pages = {20190379}	
}

@article{Barocas1997,
	title = {An Anisotropic Biphasic Theory of Tissue-Equivalent Mechanics: The Interplay Among Cell Traction,  Fibrillar Network Deformation,  Fibril Alignment,  and Cell Contact Guidance},
	volume = {119},
	ISSN = {1528-8951},
	url = {http://dx.doi.org/10.1115/1.2796072},
	DOI = {10.1115/1.2796072},
	number = {2},
	journal = {Journal of Biomechanical Engineering},
	publisher = {ASME International},
	author = {Barocas,  V. H. and Tranquillo,  R. T.},
	year = {1997},
	pages = {137-145}
}

@inbook{Barocas1994,
	title = {Biphasic Theory and In Vitro Assays of Cell-Fibril Mechanical Interactions in Tissue-Equivalent Gels},
	ISBN = {9781461384250},
	url = {http://dx.doi.org/10.1007/978-1-4613-8425-0_12},
	DOI = {10.1007/978-1-4613-8425-0_12},
	booktitle = {Cell Mechanics and Cellular Engineering},
	publisher = {Springer New York},
	author = {Barocas,  V. H. and Tranquillo,  R. T.},
	year = {1994},
	pages = {185-209}
}

@article{Lemon2006,
	title = {Mathematical modelling of engineered tissue growth using a multiphase porous flow mixture theory},
	volume = {52},
	ISSN = {1432-1416},
	url = {http://dx.doi.org/10.1007/s00285-005-0363-1},
	DOI = {10.1007/s00285-005-0363-1},
	number = {5},
	journal = {Journal of Mathematical Biology},
	publisher = {Springer Science and Business Media LLC},
	author = {Lemon,  Greg and King,  John R. and Byrne,  Helen M. and Jensen,  Oliver E. and Shakesheff,  Kevin M.},
	year = {2006},
	pages = {571–594}
}

@article{Pohlmeyer2013,
	title = {Cyclic Loading of Growing Tissue in a Bioreactor: Mathematical Model and Asymptotic Analysis},
	volume = {75},
	ISSN = {1522-9602},
	url = {http://dx.doi.org/10.1007/s11538-013-9902-x},
	DOI = {10.1007/s11538-013-9902-x},
	number = {12},
	journal = {Bulletin of Mathematical Biology},
	publisher = {Springer Science and Business Media LLC},
	author = {Pohlmeyer,  J. V. and Cummings,  L. J.},
	year = {2013},
	pages = {2450–2473}
}

@article{HOLDEN2019,
	title = {A multiphase multiscale model for nutrient-limited tissue growth,  part II: a simplified description},
	volume = {61},
	ISSN = {1446-8735},
	url = {http://dx.doi.org/10.1017/s1446181119000130},
	DOI = {10.1017/s1446181119000130},
	number = {4},
	journal = {The ANZIAM Journal},
	publisher = {Cambridge University Press (CUP)},
	author = {Holden,  E. C. and Chapman,  S. J. and Brook,  B. S. and O’Dea,  R. D.},
	year = {2019},
	pages = {368–381}
}

@article{ODea2014,
	title = {A multiscale analysis of nutrient transport and biological tissue growthin vitro},
	volume = {32},
	ISSN = {1477-8602},
	url = {http://dx.doi.org/10.1093/imammb/dqu015},
	DOI = {10.1093/imammb/dqu015},
	number = {3},
	journal = {Mathematical Medicine and Biology},
	publisher = {Oxford University Press (OUP)},
	author = {O’Dea,  R. D. and Nelson,  M. R. and El Haj,  A. J. and Waters,  S. L. and Byrne,  H. M.},
	year = {2014},
	pages = {345–366}
}

@article{Borgiani2017,
	title = {Multiscale Modeling of Bone Healing: Toward a Systems Biology Approach},
	volume = {8},
	ISSN = {1664-042X},
	url = {http://dx.doi.org/10.3389/fphys.2017.00287},
	DOI = {10.3389/fphys.2017.00287},
	journal = {Frontiers in Physiology},
	publisher = {Frontiers Media SA},
	author = {Borgiani,  Edoardo and Duda,  Georg N. and Checa,  Sara},
	year = {2017},
}

@article{C-MMS,
	title={On a mathematical model for tissue regeneration}, 
	author={Shimi Chettiparambil Mohanan and Nishith Mohan and Christina Surulescu},
	year={2024},
	journal={arXiv:2403.04516},
	DOI={10.48550/arXiv.2403.04516},
}

\end{document}